\documentclass[acmsmall]{acmart}
\usepackage{subfigure}
\usepackage{lscape}
\usepackage{threeparttable}
\usepackage{multirow}
\usepackage{array}
\newcommand{\PreserveBackslash}[1]{\let\temp=\\#1\let\\=\temp}
\newcolumntype{C}[1]{>{\PreserveBackslash\centering}p{#1}}
\newcolumntype{R}[1]{>{\PreserveBackslash\raggedleft}p{#1}}
\AtBeginDocument{%
  \providecommand\BibTeX{{%
    \normalfont B\kern-0.5em{\scshape i\kern-0.25em b}\kern-0.8em\TeX}}}

\setcopyright{acmcopyright}
\copyrightyear{2022}
\acmYear{2022}
\acmDOI{123456.123456}

\acmJournal{TOMM}
\acmVolume{10}
\acmNumber{10}
\acmArticle{10}
\acmMonth{10}

\begin{document}

\title{Deep Learning-Based Intra Mode Derivation for Versatile Video Coding}

\author{Linwei Zhu}
\email{lw.zhu@siat.ac.cn}
\author{Yun Zhang}
\email{yun.zhang@siat.ac.cn}
\author{Na Li}
\email{na.li1@siat.ac.cn}
\affiliation{%
  \institution{Shenzhen Institute of Advanced Technology, Chinese Academy of Sciences}
  \city{Shenzhen}
  \state{Guangdong}
  \country{China}
  \postcode{518055}
}

\author{Gangyi Jiang}
\email{jianggangyi@nbu.edu.cn}
\affiliation{%
  \institution{Faculty of Information Science and Engineering, Ningbo University}
  \city{Ningbo}
  \state{Zhejiang}
  \country{China}
  \postcode{315211}
  }

\author{Sam Kwong}
\email{cssamk@cityu.edu.hk}
\affiliation{%
  \institution{Department of Computer Science, City University of Hong Kong}
  \city{Hong Kong}
  \country{China}
}

\renewcommand{\shortauthors}{L. Zhu et al.}

\begin{abstract}
  In intra coding, Rate Distortion Optimization (RDO) is performed to achieve the optimal intra mode from a pre-defined candidate list. The optimal intra mode is also required to be encoded and transmitted to the decoder side besides the residual signal, where lots of coding bits are consumed. To further improve the performance of intra coding in Versatile Video Coding (VVC), an intelligent intra mode derivation method is proposed in this paper, termed as Deep Learning based Intra Mode Derivation (DLIMD). In specific, the process of intra mode derivation is formulated as a multi-class classification task, which aims to skip the module of intra mode signaling for coding bits reduction. The architecture of DLIMD is developed to adapt to different quantization parameter settings and variable coding blocks including non-square ones, which are handled by one single trained model. Different from the existing deep learning based classification problems, the hand-crafted features are also fed into the intra mode derivation network besides the learned features from feature learning network. To compete with traditional method, one additional binary flag is utilized in the video codec to indicate the selected scheme with RDO. Extensive experimental results reveal that the proposed method can achieve 2.28\%, 1.74\%, and 2.18\% bit rate reduction on average for Y, U, and V components on the platform of VVC test model, which outperforms the state-of-the-art works.
\end{abstract}



\maketitle

\section{Introduction}
With the rapid development of information technology, videos have been applied to the fields of entertainment, surveillance, education, and so on. To adapt to more applications in our daily life, the videos have evolved in various dimensions in the last decade, including \textbf{High Definition} (\textbf{HD}), \textbf{Wide Color Gamut} (\textbf{WCG}) \cite{7174544}, \textbf{High Dynamic Range} (\textbf{HDR}) \cite{7174544}, \textbf{Multi-view Video plus Depth} (\textbf{MVD}) \cite{6519266}, 360 degree video \cite{8902161}, light field image/video \cite{8030107}, and dynamic point cloud \cite{8571288}. Unfortunately, from low dimension to high dimension, the dramatically increased video data challenges the limited storage space and transmission bandwidth. From H.264/\textbf{Advanced Video Coding} (\textbf{AVC}) \cite{1218189}, \textbf{High Efficiency Video Coding} (\textbf{HEVC}) \cite{6316136}, to the state-of-the-art \textbf{Versatile Video Coding} (\textbf{VVC}) \cite{9328514} that was issued in 2020, although a large compression ratio has been achieved, it still cannot catch up with the increase of video data. Advanced video compression algorithm is always desired to maximize the visual quality at a given bandwidth budget. \par

In the framework of existing hybrid video coding, the modules mainly consist of intra/inter prediction, transform, quantization, entropy encoding and in-loop filtering. To improve compression efficiency, a variety of novel coding tools have been developed in the issued standards, including \textbf{QuadTree plus Multi-Type Tree} (\textbf{QT}+\textbf{MTT}) structure \cite{8859290} for coding block partition, \textbf{Matrix-based Intra Prediction} (\textbf{MIP}) \cite{9190883} and \textbf{Cross-Component Linear Model} (\textbf{CCLM}) \cite{8350031} for intra luma and chroma prediction, \textbf{History-based Motion Vector Prediction} (\textbf{HMVP}) \cite{8712795} and \textbf{Decoder-side MV Refinement} (\textbf{DMVR}) \cite{9252910} for motion estimation/compensation, \textbf{Multiple Transform Selection} (\textbf{MTS}) \cite{6616629} for transform, CABAC engine with multi-hypothesis probability estimation for entropy encoding, \textbf{Sample Adaptive Offset} (\textbf{SAO}) and \textbf{Adaptive Loop Filter} (\textbf{ALF}) for in-loop filtering. These mentioned coding tools have achieved significant coding gains. \par

One of the most important modules is intra prediction \cite{6317153}, which aims to remove spatial redundancy as much as possible. Parts of the available neighboring blocks are weighted to produce the predicted block. Traditionally, intra modes include Planar, DC, and angular modes. To achieve more accurate prediction result, various algorithms have been developed. In \cite{6905757}, intra prediction was analyzed in frequency domain, and the frequency components were selectively discarded to improve the performance. Li et al. \cite{8712738} presented a bi-intra prediction method based on the binary combination of existing uni-intra prediction modes. Rather than regular out-block reference pixels, the in-block ones were employed in \cite{8470962} to perform intra prediction for screen content, and an additional in-loop residual signal was used. An iterative filtering method was employed for intra prediction in addition to the traditional intra prediction in \cite{7479465}. To achieve more reference pixels, the multi-line based scheme was presented in \cite{7762080}, where six more lines of pixels located at the above and left neighbors were collected. Different from fixed scan order, an adaptive block coding order \cite{7331144} was proposed for intra prediction to better exploit spatial correlations. In analogous to motion estimation in inter coding, \textbf{Intra Block Copy} (\textbf{IBC}) \cite{7547947} was introduced for screen content, which aims to exploit long distance correlations in an image. Two modes with high probability from gradient histogram were combined to generate a new intra mode in \cite{9102799}. In \cite{8115239}, the local and nonlocal correlations were exploited for hybrid intra prediction, where the adaptive template matching prediction, combined local and nonlocal prediction, combined neighboring modes prediction were performed. These methods mentioned above exploit spatial redundancy from neighbors with manually designed functions, which may limit the performance. Advanced schemes are desired to adapt to diverse video contents.\par

To further improve compression efficiency of intra coding, the problem of signal processing is formulated as an artificial intelligence task, where powerful neural network is adopted \cite{8693636}\cite{ZHANG2020395} and a training database for deep video compression is provided in \cite{9527119}.
In specific, the problem of intra luma prediction was formulated as an inpainting task \cite{8744274}, and the problem of intra chroma prediction was modeled as a colorization task \cite{9247080}\cite{3434250}. An iterative training strategy for neural network was presented in \cite{9266121}, where training blocks were collected from previous iteration to further improve performance. Wang et al. \cite{8794555} proposed a multi-scale convolutional network based intra prediction approach, in which the neighboring reconstructed L-shape was fed to network as well as the traditional angular intra prediction result to make a more accurate prediction. With conditional autoencoder \cite{9244548}, multi-mode intra prediction was performed for luma and chroma components. Sun et al. \cite{8947942} proposed two enhanced intra prediction schemes with multiple neural networks, where the appending scheme was to replace the traditional modes and the substitution scheme was to replace the highest and lowest probable traditional modes. In \cite{8727931}, a progressive spatial recurrent neural network was presented for intra prediction, which was able to produce prediction by passing information along from previous output.
From these existing learning based methods, it can be observed that they mostly aim to make more accurate luma and chroma predictions from regression perspective to achieve coding gains, while the module of intra mode derivation has not been exploited from classification perspective with deep learning tools.\par

In intra coding, the intra mode is also required to be encoded and transmitted to the decoder side besides residual signal. For intra mode signaling, \textbf{Most Probable Mode} (\textbf{MPM}) list, which is constructed from the neighboring blocks, plays an important role and saves significant coding bits. In \cite{8351605}, two MPM construction methods were presented for VVC, where one was extended from HEVC, and the other was sorted according to the probability of each candidate. Besides the nearest neighboring lines, Chang et al. \cite{8712640} extended MPM mechanism to \textbf{Multi-Reference Line} (\textbf{MRL}) scheme for better performance. A conditional random field model was established to re-construct the MPM list in \cite{9115218}, where the short and long range correlations were considered. In addition, a decision tree was utilized to exploit multiple dynamic lists of intra mode signaling \cite{8712638}. By investigating the occurrences of intra modes in the neighboring blocks, \textbf{Most Frequent Mode} (\textbf{MFM}) list \cite{2018.7452} was derived to compete with the existing MPM list. To skip intra mode signaling and save coding bits, Xu et al. \cite{6213253} proposed a predictive coding scheme, in which the angular correlation in spatial domain was calculated with modulo-N arithmetic operations. Additionally, template based \cite{7906340}, histogram of gradients based \cite{8712810}, and texture analysis based \cite{8803773} intra mode derivation methods were presented in a manual manner. Basically, the MPM list construction and intra mode derivation have been investigated by traditional statistics and experience, which still can be further improved with advanced learning based schemes. \par

In this work, to skip the module of intra mode signaling and save coding bits, the process of intra mode derivation is formulated as a multi-class classification task. The main contributions of this work are listed as follows. \par

\begin{enumerate}
\item The process of intra mode derivation in intra coding is modeled as a multi-class classification task, termed as \textbf{Deep Learning based Intra Mode Derivation} (\textbf{DLIMD}), which is used to skip the module of intra mode signaling for saving bits.
\item In DLIMD, the learned features and hand-crafted features are combined together for intra mode derivation. Additionally, the proposed DLIMD can be applied to variable coding blocks (including non-square blocks) and any different \textbf{Quantization Parameter} (\textbf{QP}) settings.
\item To further improve the performance, one additional binary flag is utilized to indicate the finally selected scheme from \textbf{Rate Distortion} (\textbf{RD}) cost competition. The proposed method achieves superior performance when compared with the state-of-the-art algorithms.
\end{enumerate}

The remainder of this work is organized as follows. Motivation is presented in Section 2. The proposed DLIMD for video coding is discussed in detail in Section 3. The experiments are conducted and the results are analyzed in Section 4. Section 5 concludes this work.

\begin{figure}[t]
  \centering
    \includegraphics[width=0.6\textwidth]{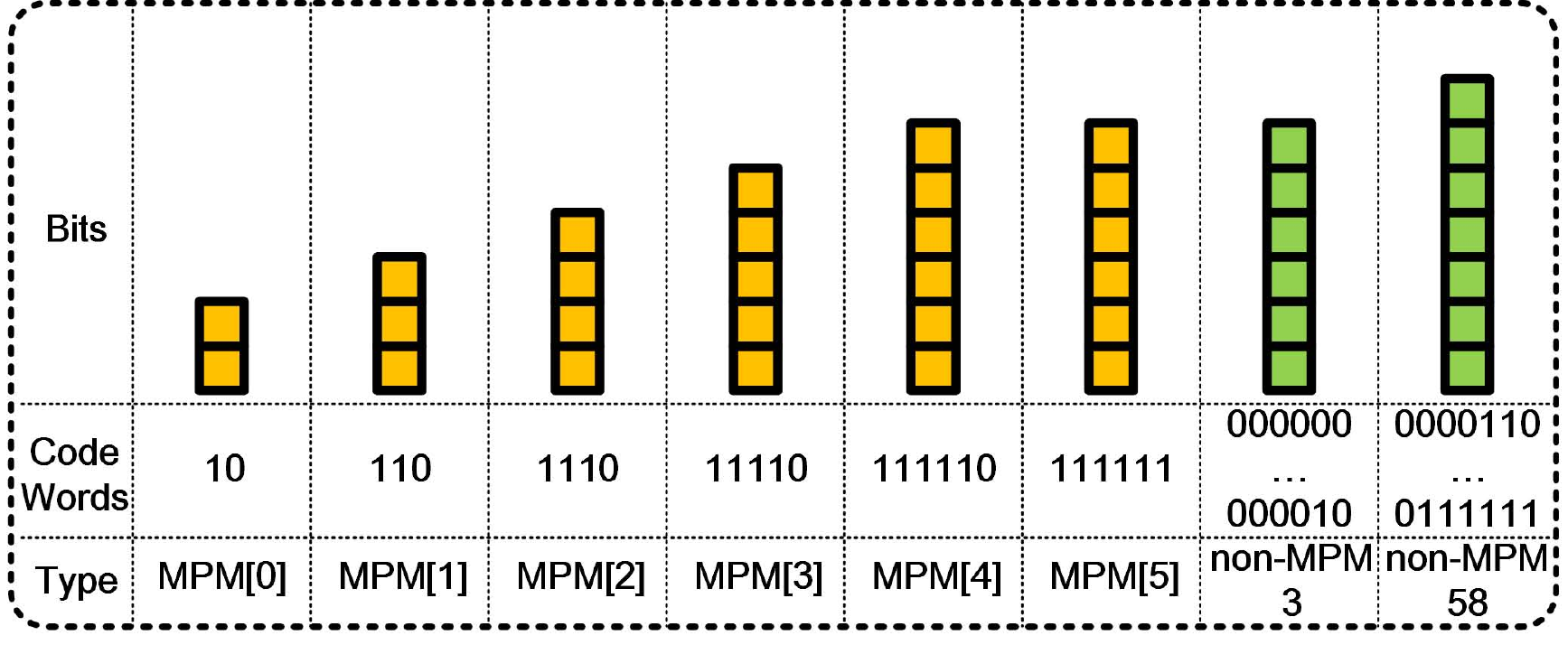}\\
  \caption{67 intra mode signaling in VVC.}
  \label{Fig4} 
\end{figure}

\begin{table*}[t]\caption{Statistical results of intra mode signaling.} \label{table1}
\footnotesize
\begin{center}
    \begin{tabular}{|c|c|C{0.9cm}|C{0.9cm}|C{0.9cm}|C{0.9cm}|C{0.9cm}|C{0.9cm}|C{0.9cm}|C{0.9cm}|}
    \cline {1-10} \multirow{3}{*}{Class}&\multirow{3}{*}{Sequence}&\multicolumn{4}{c|}{Coding bits per intra mode (BPM)}&\multicolumn{4}{c|}{Percentage of coding bits of intra mode}\\
                  &&\multicolumn{4}{c|}{$\alpha$}&\multicolumn{4}{c|}{$\beta$}\\
    \cline {3-10} &&{QP=22}&{QP=27}&{QP=32}&{QP=37}&{QP=22}&{QP=27}&{QP=32}&{QP=37}\\
    \hline
    \hline
               \multirow{2}{*}{A}   &Tango2&2.18&2.90&2.92&2.97&5.39\%&10.9\%&14.7\%&18.3\% \\
    \cline{2-10}                     &FoodMarket4&2.45&2.61&2.63&2.71&7.14\%&10.4\%&12.9\%&14.9\% \\
    \cline{1-10}\multirow{2}{*}{B}   &BasketballDrive&2.94&2.92&2.93&2.85&6.60\%&11.5\%&15.6\%&18.3\% \\
    \cline{2-10}                     &BQTerrace&3.40&3.57&3.53&3.44&5.27\%&9.46\%&14.1\%&18.9\% \\
    \cline{1-10}\multirow{2}{*}{C}   &BQMall&3.84&3.86&3.75&3.65&8.95\%&12.2\%&16.3\%&20.8\% \\
    \cline{2-10}                     &BasketballDrill&3.52&3.55&3.58&3.76&14.6\%&18.4\%&21.0\%&25.1\% \\
    \cline{1-10}\multirow{2}{*}{D}   &BlowingBubbles&4.34&4.25&4.17&3.82&7.88\%&11.5\%&16.3\%&20.6\% \\
    \cline{2-10}                     &BasketballPass&3.85&4.04&3.93&3.67&8.82\%&11.8\%&16.4\%&22.1\% \\
    \cline{1-10}\multirow{2}{*}{E}   &FourPeople&3.59&3.62&3.57&3.53&9.80\%&13.8\%&17.7\%&21.8\% \\
    \cline{2-10}                     &Johnny&3.37&3.43&3.43&3.50&8.39\%&13.1\%&18.7\%&23.2\% \\
    \cline{1-10}
    \multicolumn{2}{|c|}{{AVERAGE}}&{3.35}&{3.48}&{3.44}&{3.39}&8.28\%&12.3\%&16.4\%&20.4\%\\
    \hline
   \end{tabular}
\end{center}
\end{table*}

\section{Motivation}
In VVC, intra coding modes/tools  \cite{9400392} include DC, Planar, 65 angular modes, \textbf{Wide Angle Intra Prediction} (\textbf{WAIP}), MRL, \textbf{Position Dependent Prediction Combination} (\textbf{PDPC}), MIP, \textbf{Intra-Sub Partition} (\textbf{ISP}), and CCLM. It should be mentioned that the intra mode is also required to be encoded and transmitted to the decoder side. To effectively signal these intra modes to the decoder side, the derivation is performed with intra modes from neighbors, where six of them are produced and accommodated to the MPM list. Generally, the first one in the MPM list is always fixed, i.e., Planar mode, which is encoded with two-bit length. The other five MPMs are achieved according to spatial correlation from the neighbors, and encoded with three-bit to six-bit length. The non-MPM modes are divided into two parts which contain 3 and 58 modes, respectively. They are truncated binary coded with six-bit and seven-bit length. The detailed intra modes signaling can be found in Fig. \ref{Fig4}. In addition, statistical experiments are conducted under the platform of \textbf{VVC Test Model} version 5.0 (\textbf{VTM} 5.0) to present coding \textbf{Bits Per intra Mode} (\textbf{BPM}), where ten sequences with various contents from different classes are encoded under \textbf{All Intra} (\textbf{AI}) configuration. The value of BPM is calculated by the total coding bits of intra mode against the number of intra blocks, where the coding bits are collected after CABAC entropy encoding. The statistical results are shown in the left columns of Table \ref{table1} and the values of BPM are 3.35, 3.48, 3.44, and 3.39 on average under four QP settings. \par

Furthermore, to demonstrate how many bits are spent in the module of intra mode signaling, the percentage of coding bits of intra mode in a frame is collected, and illustrated in the right columns of Table \ref{table1}. It can be found that this percentage increases from 8.28\% to 20.4\% on average as QP value increases. In the case of small QP settings, the percentage is limited, because the coding bits of residue (the difference between prediction and source) are much larger than those of intra mode, while in the case of large QP settings, the coding bits of residue become limited, which results in high percentage of coding bits of intra mode. From these results, we can conclude that if more advanced intra mode signaling approach is presented, the coding performance can be further improved. \par

\section{Proposed Deep Learning based Intra Mode Derivation for Video Coding}
\subsection{Problem Formulation and Framework}
In this work, we focus on the optimization of DC, Planar, and 65 angular modes signaling. According to Fig. \ref{Fig4}, the straightforward idea of improving coding performance is to predict the best intra mode from all 67 candidates and place it to the first in MPM list. This intra mode derivation can achieve promising performance, because the intra mode signaling only consumes two bits, which is less than other cases. However, it still can be further improved by skipping the RD checking process and intra mode signaling to save coding bits.\par

The optimal intra mode of current block is finally selected based on the minimum RD cost by checking the candidate list in VTM. This process can be represented by the following equation,
\begin{equation}
n^* = \mathop{\arg\min}_{n}\{D_n + \lambda (R_n^{r} + R_n^{m} + R_n^{o})\},
\end{equation}
where $n$ indicates the index of intra mode, $n \in [0, 66]$ for Planar, DC, and 65 angular modes, $D_n$ is the distortion, $\lambda$ is the Lagrange Multiplier, $R_n^r$, $R_n^m$, and $R_n^o$ indicate the coding bits of residue, intra mode, and other information, respectively.

\begin{figure*}[t]
  \centering
    \includegraphics[width=0.7\textwidth]{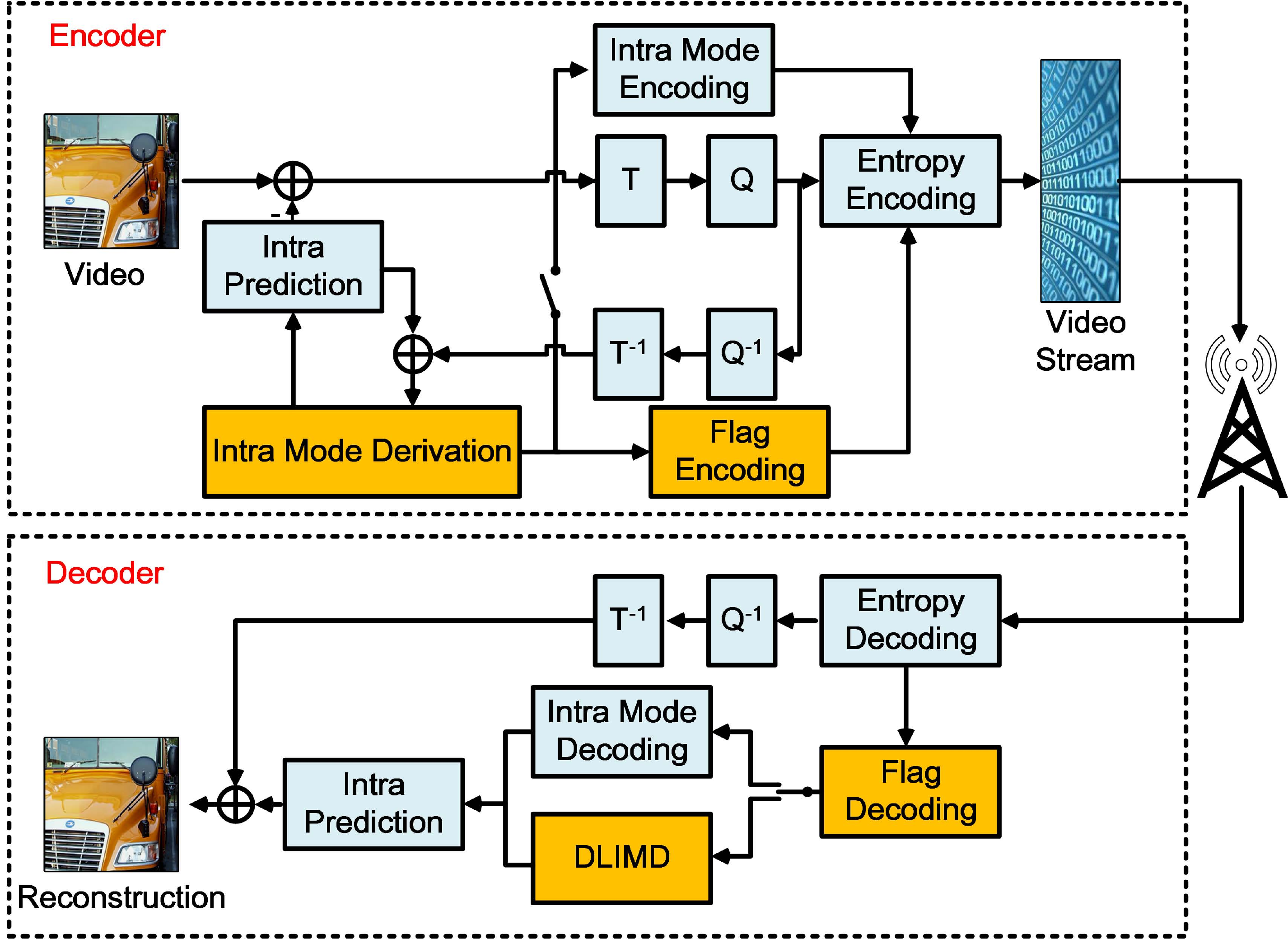}\\
  \caption{Framework of proposed deep learning based intra mode derivation for video coding.}
  \label{Fig7} 
\end{figure*}

According to Eq. (1), to achieve the optimal one from a pre-defined candidate list, this process can be formulated as a multi-class classification task. Generally, the construction of MPM list can be regarded as a manual classification scheme, and top-6 intra modes are manually selected. To further improve the performance, we aim to solve this multi-class classification task with a deep learning approach. In specific, the optimal intra mode can be derived directly instead of checking candidate list, and the module of intra mode signaling is expected to be skipped for coding bits reduction.
Fig. \ref{Fig7} illustrates the framework of proposed deep learning based intra mode derivation for video coding. T and Q indicate transform and quantization, T$^{-1}$ and Q$^{-1}$ indicate inverse transform and inverse quantization.

In the video encoder, intra mode derivation is performed, including the conventional intra mode check and the proposed DLIMD. According to RD cost, only one of them will be finally selected. If DLIMD is selected, the strategy flag is set as 1 and the switch is opened for skipping the module of intra mode encoding; otherwise, the strategy flag is set as 0 and the switch is closed for activating the module of intra mode encoding. The strategy flag is always encoded and transmitted to indicate the selected scheme. It is worth mentioning that the other modules in video codec are not changed. The RD cost competition between DLIMD and traditional method (including DC, Planar, angular modes, MRL, MIP and ISP) can be represented by the following equation.
\begin{equation}
S^* = \mathop{\arg\min}_{S}\{D_S + \lambda (R_S^{r} + R_S^{m} + R_S^{f} + R_S^{o})\},
\end{equation}
where $S^*$ indicates the selected scheme, i.e., $S\in$ \{DLIMD, traditional method\}, $D_S$ is the distortion under $S$ scheme, $\lambda$ is the Lagrange Multiplier, $R_S^r$, $R_S^m$, $R_S^f$, and $R_S^o$ indicate the coding bits of residue, intra mode, strategy flag, and other information, respectively. In addition, it should be noted that if the proposed DLIMD is selected, there is no coding bit for intra mode, i.e., $R_S^m = 0$.

In the video decoder, the strategy flag is firstly decoded before intra prediction. If this strategy flag is 0, the intra mode will be decoded directly; otherwise, the intra mode will be derived by the proposed DLIMD. With the intra mode, intra prediction is performed accordingly. Finally, the prediction result plus decoded residual information produces the reconstruction. \par


To estimate the upper bound of performance under the proposed framework, we define that $\alpha$ and $\beta$ are the original value of BPM and the percentage of coding bits of intra mode in a frame, the statistical values of them are illustrated in Table \ref{table1}, $\gamma$ is the percentage of selected intra blocks under the proposed scheme. One additional binary flag is utilized for indication between the proposed scheme and the original scheme, which is encoded by context mode. Then, the value of BPM becomes $- \gamma \times log_2(\gamma) + (1 - \gamma)\times{(\alpha - log_2(1 - \gamma)})$. Accordingly, the bit saving can be calculated as follows,
\begin{equation}
\eta = \frac{\alpha - [- \gamma \times log_2(\gamma) + (1 - \gamma)\times{(\alpha - log_2(1 - \gamma)})]}{\alpha} \times \beta.
\end{equation}
The condition of upper bound is $\gamma = 100\%$, then the bit saving equals to $\beta$. As such, the upper bound of bit saving can reach 8.28\%, 12.3\%, 16.4\%, and 20.4\% when QP value equals to \{22, 27, 32, 37\}, respectively.

\begin{figure*}[t]
  \centering
    \includegraphics[width=0.98\textwidth]{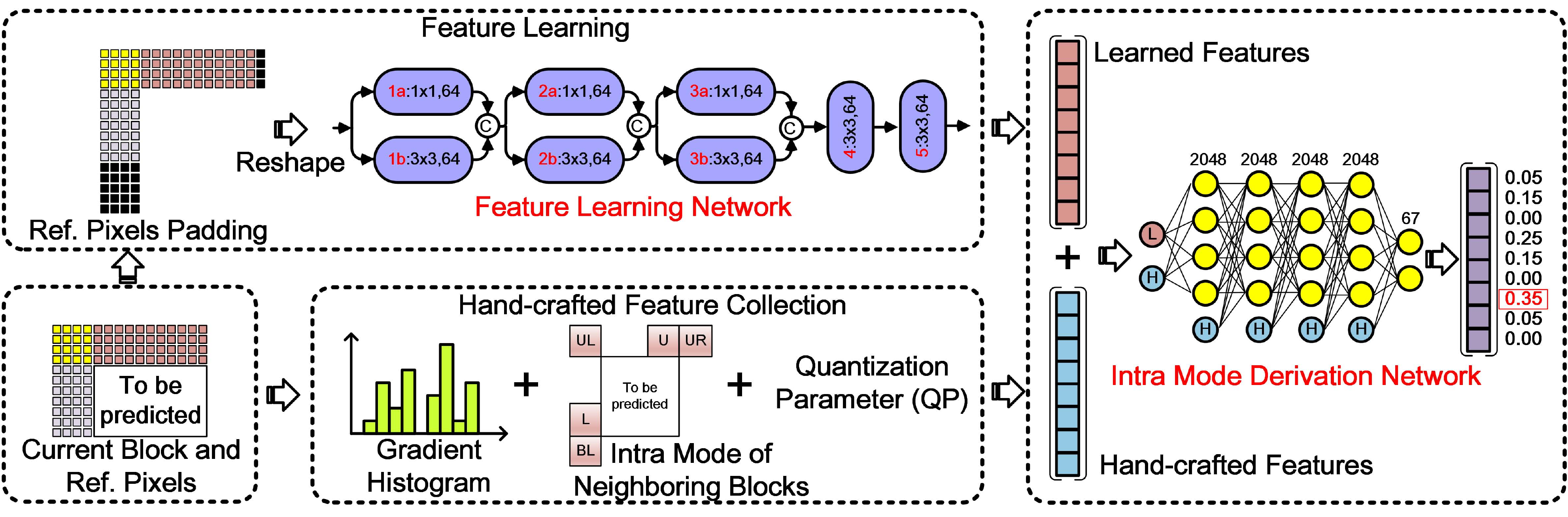}\\
  \caption{Architecture of deep learning based intra mode derivation.}
  \label{Fig6} 
\end{figure*}

\subsection{Deep Learning based Intra Mode Derivation}
Fig. \ref{Fig6} illustrates the proposed architecture of deep learning based intra mode derivation scheme, in which two neural networks are included, one is feature learning network and the other is intra mode derivation network. The former is used to extract the highly dimensional features and the latter aims to infer the optimal intra mode directly without RD cost checking. In particular, the hand-crafted and learned features are supposed to be combined to enjoy their individual benefits. The detailed hyper parameters of these two networks are listed in Tables \ref{table2} and \ref{table3}. \par

In the feature learning network, five convolutional layers are included, and the first three ones (each has two sub-layers) are placed in a parallel manner. The kernel sizes are $1\times1$ and $3\times3$ in convolutional layers. \textbf{Rectified Linear Unit} (\textbf{ReLU}) is employed as activation function. The number of feature maps in every convolutional layer is 64. In the intra mode derivation network, five fully connected layers are included, and the node of each layer except the last one is 2048. In the last fully connected layer, the activation function is SoftMax, and the number of nodes becomes 67, which aims to match the number of intra modes. For the input of every fully connected layer, it is always combined with the hand-crafted features, which is represented by,
\begin{equation}
\textbf{I}_i^f = {\rm{concat}}(\textbf{O}^f_{i-1}, \textbf{f}_0), i \in [1,5],
\end{equation}
where $\textbf{O}^f_{i-1}$ is the output of the $(i-1)^{th}$ fully connected layer, $\textbf{f}_0$ indicates the hand-crafted features, $\textbf{O}^f_0$ is the reshaped vector of output of feature learning network. To avoid overfitting, the dropout is performed at every fully connected layer. The dropout rate is set as 0.5 at the training stage, and set as 1.0 at the testing stage.

\begin{table}[t]\caption{Hyper-parameters of the feature learning network.}\label{table2}
\footnotesize
\begin{center}
    \begin{tabular}{|c|c|c|c|c|c|}
    \cline {1-6} {\#}&{Type}&{Kernel}&{Stride}&{Outputs}&{Activation}\\
    \cline{1-6}
    \hline
    \hline
                                        1a&\multirow{8}{*}{CNN}&{$1\times1$}&\multirow{8}{*}{1}&\multirow{8}{*}{64}&\multirow{8}{*}{ReLU}\\
    \cline{1-1}\cline{3-3}   1b&&{$3\times3$}&&&\\
    \cline{1-1}\cline{3-3}   2a&&{$1\times1$}&&&\\
    \cline{1-1}\cline{3-3}   2b&&{$3\times3$}&&&\\
    \cline{1-1}\cline{3-3}   3a&&{$1\times1$}&&&\\
    \cline{1-1}\cline{3-3}   3b&&{$3\times3$}&&&\\
    \cline{1-1}\cline{3-3}   4&&{$3\times3$}&&&\\
    \cline{1-1}\cline{3-3}   5&&{$3\times3$}&&&\\
    \cline{1-6}
   \end{tabular}
\end{center}
\end{table}

\begin{table}[t]\caption{Hyper-parameters of the intra mode derivation network.}\label{table3}
\footnotesize
\begin{center}
    \begin{tabular}{|c|c|c|c|c|}
    \cline {1-5} {\#}&{Type}&{Input Size}&{Nodes}&{Activation}\\
    \cline{1-5}
    \hline
    \hline
                                       1&\multirow{5}{*}{FCN}&{33792+73}&\multirow{4}{*}{2048}&\multirow{4}{*}{ReLU}\\
    \cline{1-1}\cline{3-3}  2&&\multirow{4}{*}{2048+73}&&\\
    \cline{1-1}                        3&&&&\\
    \cline{1-1}                        4&&&&\\
    \cline{1-1}\cline{4-5}             5&&&{67}&SoftMax\\
    \cline{1-5}
   \end{tabular}
\end{center}
\end{table}

With the neighboring blocks and reference pixels, 73 hand-crafted features are collected, including sixty-seven features from gradient histogram, five features from intra mode of neighboring blocks, and QP value. The gradient histogram can be regarded as the probability of each candidate intra mode, and its detailed calculation can be found in \cite{8803773}. Due to the highly spatial correlation, the \textbf{Up-Left} (\textbf{UL}), \textbf{Up} (\textbf{U}), \textbf{Up-Right} (\textbf{UR}), \textbf{Left} (\textbf{L}) and \textbf{Bottom Left} (\textbf{BL}) blocks are used to provide their final selected intra modes as five hand-crafted features, as shown in the module of hand-crafted feature collection in Fig. \ref{Fig6}. The QP value balances the reconstruction quality and coding bits, i.e., lower QP value indicates better reconstruction quality and more coding bits, and vice versa, which has an impact on the intra mode derivation. In addition, with the feature of QP value, there is no need to train different networks for different QP settings. In the case of boundary, the intra modes of neighbors are unavailable, which are initially set as Planar mode. \par

Due to lossy video coding, the neighboring blocks and reference pixels are degraded, which may affect the hand-crafted features, especially for the gradient histogram calculation. Therefore, the learned features are employed. As mentioned before, the coding block is not fixed, which can be flexibly partitioned from $128\times128$ to $4\times4$, including non-square patterns. Accordingly, the number of reference pixels is different for variable coding blocks. For luma coding block, it follows that the width and height belong to \{4, 8, 16, 32, 64\} \cite{9400392}. Therefore, it seems that 25 networks are required, which challenges computational and storage resources. It is expected that one single trained network can be applied to variable coding blocks. In the designed architecture, the multi-line reference pixels are collected and padded to a fixed size to adapt to variable coding blocks, where the fixed memory is allocated under the maximum available coding block $64 \times 64$. Suppose the current block has a dimension of $4\times4$, the padding is performed to the lines that exceed the current block size of $4\times4$, where 60 padded lines from left and 60 padded lines from top are always used. A matrix with size of $(64 + 4 + 64)\times4$ is fed to the feature learning network. Finally, the learned features with size of $(64 + 4 + 64)\times 4\times 64 = 33792$ are produced and reshaped to one dimension of vector. \par

In addition, the number of \textbf{FLoating-point OPerations} (\textbf{FLOPs}) \cite{MolchanovTKAK17} is used to evaluate the complexity of neural network. For convolutional layer,
\begin{equation}
FLOPs = 2H \times W \times(C_{in}\times K^2 + 1) \times C_{out},
\end{equation}
where $H$, $W$ and $C_{in}$ are height, width and number of channels of the input feature map, $K$ is the kernel size, and $C_{out}$ is the number of output channels. The values of FLOPs in convolutional layers are $1.3\times10^5$, $6.7\times10^5$, $8.7\times10^6$, $7.7\times10^7$, $8.7\times10^6$, $7.7\times10^7$, $7.7\times10^7$, and $3.9\times10^7$, respectively. For fully connected layer,
\begin{equation}
FLOPs = (2I - 1)\times O,
\end{equation}
where $I$ is the input dimensionality and $O$ is the output dimensionality. The values of FLOPs in fully connected layers are $1.3\times10^8$, $8.6\times10^6$, $8.6\times10^6$, $8.6\times10^6$, and $2.8\times10^5$, respectively.

\subsection{Neural Network Training}
The DIV2K database \cite{8014883} with 900 images (the resolution changes from $2040\times648$ to $2040\times2040$) is used to generate the neural network training set. These images are all resized to $2048\times1536$ and packed as a sequence from RGB color space to YCbCr color space. Then, this pseudo sequence is encoded by VTM 5.0 under default AI configuration with QPs \{22, 27, 32, 37\} to collect the training samples. During the process of video coding, the hand-crafted features and source reference pixels of current block are collected with the associated label (intra mode) regardless of the coding block size. According to Fig. 3 shown in \cite{9301752}, the distribution of intra modes is unequal. The Planar, DC, horizontal, and vertical modes are more frequently selected than other intra modes. Additionally, the unbalanced data may make the multi-class classification network training failure. Thereby, the number of training samples for each label and QP is fixed as 50000, and then the total number is $50000 \times 4 \times 67$. Totally, the volume of training data reaches about 80 GB. In addition, for the purpose of validation during training, $20000\times67$ samples are selected in the training set, where the number of validation samples is 20000 for each label. \par

In this work, the Tensorflow package is adopted for network training on NVIDA GeForce 1080 Ti GPU with AdamOptimizer. The memory of workstation is 112G, which is able to accommodate the training samples.  For this multi-class classification task, the cross entropy is utilized as the loss function,
\begin{equation}
L = -\frac{1}{N}\sum^N_{i=1} \sum^M_{j=1} \{y_{j}^{i} \times ln (x_{j}^{i})\},
\end{equation}
where $N$ is the number of training samples in a batch, $M$ is the number of classes, i.e., $M=67$, $y_{j}^{i}$ is the ground truth of the $i^{th}$ training sample, and $x_{j}^{i}$ is the output of intra mode derivation network after softmax layer. It should be noted that the ground truth is represented in the one-hot manner. For example, if the intra mode is 4 for the $i^{th}$ training sample, $y^i_4 = 1$ and $y^i_j = 0$ $(j \neq 4)$, it also can be rewritten as $\textbf{y}^i = [y^i_1, y^i_2, \dots, y^i_j, \dots, y^i_{67}] = [0, 0, 0, 1, 0, \dots, 0]$. The batch size and number of epochs are set as 1024 and 1000. The initial learning rate $r_0$ is $1\times10^{-4}$, and it is always updated after each epoch, i.e., $r_0 \times 0.999^i$, where $i$ is the index of training epoch.

\section{Experimental Results and Analyses}

\begin{table*}[t]\caption{Coding bits per intra mode under the proposed method.} \label{table5}
\footnotesize
\begin{center}
    \begin{tabular}{|c|c|c|c|c|c|c|c|c|c|}
    \cline {1-10} \multirow{3}{*}{Class}&\multirow{3}{*}{Sequence}&\multicolumn{4}{c|}{Current value of BPM}&\multicolumn{4}{c|}{Coding bit saving of intra mode}\\
                  &&\multicolumn{4}{c|}{$\alpha'$}&\multicolumn{4}{c|}{$\eta'$}\\
    \cline {3-10} &&{QP=22}&{QP=27}&{QP=32}&{QP=37}&{QP=22}&{QP=27}&{QP=32}&{QP=37} \\
    \hline
    \hline
                \multirow{2}{*}{A}   &Tango2&1.02&1.46&1.44&1.71&53.2\%&49.7\%&50.7\%&42.4\%\\
    \cline{2-10}                     &FoodMarket4&1.33&1.50&1.53&1.82&45.7\%&42.5\%&41.8\%&32.8\% \\
    \cline{1-10}\multirow{2}{*}{B}   &BasketballDrive&1.99&1.83&1.65&1.69&32.3\%&37.3\%&43.7\%&40.7\% \\
    \cline{2-10}                     &BQTerrace&2.76&2.53&2.36&2.25&18.8\%&29.1\%&33.1\%&34.6\%\\
    \cline{1-10}\multirow{2}{*}{C}   &BQMall&2.84&2.78&2.59&2.40&26.0\%&28.0\%&30.9\%&34.2\%\\
    \cline{2-10}                     &BasketballDrill&2.73&2.60&2.49&2.58&22.4\%&26.8\%&30.4\%&31.4\%\\
    \cline{1-10}\multirow{2}{*}{D}   &BlowingBubbles&3.09&2.92&2.71&2.45&28.8\%&31.3\%&35.0\%&35.9\%\\
    \cline{2-10}                     &BasketballPass&3.16&3.02&2.78&2.57&17.9\%&25.2\%&29.3\%&30.0\%\\
    \cline{1-10}\multirow{2}{*}{E}   &FourPeople&2.51&2.40&2.31&2.23&30.1\%&33.7\%&35.3\%&36.8\%\\
    \cline{2-10}                     &Johnny&2.41&2.45&2.34&2.32&28.5\%&28.6\%&31.8\%&33.7\% \\
    \cline{1-10}
    \multicolumn{2}{|c|}{{AVERAGE}}&{2.38}&{2.35}&{2.22}&{2.20}&{30.4\%}&{33.2\%}&{36.2\%}&{35.3\%}\\
    \hline
   \end{tabular}
\end{center}
\end{table*}

\begin{table*}[t]\caption{Performance comparison in terms of BD-BR with QPs \{22, 27, 32, 37\}. (Unit: \%)} \label{table6}
\footnotesize
\begin{center}
    \begin{tabular}{|C{0.48cm}|C{1.68cm}|C{0.55cm}|C{0.55cm}|C{0.55cm}|C{0.55cm}|C{0.55cm}|C{0.55cm}|C{0.55cm}|C{0.55cm}|C{0.55cm}|C{0.55cm}|C{0.55cm}|C{0.55cm}|}
    \cline {1-14} \multirow{2}{*}{Class}&\multirow{2}{*}{Sequence}&\multicolumn{3}{c|}{Narsallah's \cite{8803773}}&\multicolumn{3}{c|}{Abdoli's \cite{9102799}}&\multicolumn{3}{c|}{Li's \cite{9115218}}&\multicolumn{3}{c|}{Proposed}\\
    \cline {3-14} &&{Y}&{U}&{V}&{Y}&{U}&{V}&{Y}&{U}&{V}&{Y}&{U}&{V}\\
    \hline
    \hline
                \multirow{3}{*}{A1}      &Tango2&-0.20&-0.41&0.01&-0.69&-0.91&-0.16&-0.08&-0.04&0.13&-2.50&-2.77&-2.20 \\
    \cline{2-14}                         &FoodMarket4&-0.36&0.06&-0.44&-0.81&-0.33&-0.68&-0.02&-0.14&0.00&-2.65&-1.66&-1.14\\
    \cline{2-14}                         &Campfire&-0.18&-0.16&-0.34&-0.42&-0.20&-0.12&-0.08&0.00&-0.19&-2.38&-1.14&-1.98\\
    \cline{1-14}\multirow{3}{*}{A2}      &CatRobot1&-0.13&-0.29&-0.33&-0.33&-0.08&-0.36&-0.07&-0.08&-0.01&-2.69&-1.90&-2.44\\
    \cline{2-14}                         &DaylightRoad2&0.02&-0.29&0.01&-0.29&-0.32&-0.10&-0.26&-0.13&-0.15&-2.70&-2.66&-2.63\\
    \cline{2-14}                         &ParkRunning3&-0.04&-0.02&-0.11&-0.27&-0.25&-0.29&-0.05&-0.07&-0.08&-1.04&-0.85&-0.88\\
    \cline{1-14}\multirow{5}{*}{B}       &MarketPlace&-0.12&-0.08&-0.09&-0.35&0.00&-0.43&-0.10&-0.23&-0.20&-2.29&-1.51&-1.61\\
    \cline{2-14}                         &RitualDance&-0.46&-0.41&-0.33&-0.65&-0.32&-0.31&-0.02&-0.23&-0.20&-1.81&-1.67&-1.55\\
    \cline{2-14}                         &Cactus&-0.06&0.09&0.02&-0.35&0.03&-0.21&-0.07&-0.18&-0.06&-2.49&-1.17&-3.41\\
    \cline{2-14}                         &BasketballDrive&-0.20&-0.67&-0.16&-0.67&-0.82&-0.25&-0.10&-0.61&-0.35&-2.42&-2.68&-2.04\\
    \cline{2-14}                         &BQTerrace&0.00&-0.28&-0.11&-0.27&-0.37&-0.26&-0.15&-0.31&-0.41&-1.88&-1.77&-2.40\\
    \cline{1-14}\multirow{4}{*}{C}       &BasketballDrill&0.15&0.04&0.80&-0.27&0.14&0.89&-0.31&-0.01&-0.69&-2.29&1.18&-2.62\\
    \cline{2-14}                         &BQMall&-0.37&-0.50&0.14&-0.52&-0.27&-0.37&-0.01&0.40&-0.01&-2.89&-1.93&-1.36\\
    \cline{2-14}                         &PartyScene&-0.18&-0.11&-0.16&-0.35&-0.20&-0.29&-0.16&0.02&-0.03&-1.93&-2.45&-0.71\\
    \cline{2-14}                         &RaceHorsesC&-0.16&-0.16&-0.39&-0.49&-0.16&-0.11&0.01&0.08&-0.08&-1.78&-0.89&-2.01\\
    \cline{1-14}\multirow{4}{*}{D}       &BasketballPass&-0.20&-0.14&-0.02&-0.41&-0.01&-0.38&0.04&-0.50&-0.59&-1.67&-2.31&-4.17\\
    \cline{2-14}                         &BQSquare&-0.33&-0.27&-0.02&-0.23&-0.08&-0.24&-0.01&0.14&0.04&-1.92&-0.13&-1.73\\
    \cline{2-14}                         &BlowingBubbles&-0.27&-0.68&-0.95&-0.69&-0.53&-0.96&0.03&-0.55&-0.34&-2.09&-1.36&-1.67\\
    \cline{2-14}                         &RaceHorses&-0.25&0.29&0.38&-0.43&-0.65&-0.38&-0.04&-0.03&0.69&-1.84&-2.48&-1.42\\
    \cline{1-14}\multirow{3}{*}{E}       &FourPeople&-0.41&-0.52&-0.38&-0.55&-0.77&0.04&0.03&-0.17&0.00&-3.21&-2.61&-2.28\\
    \cline{2-14}                         &Johnny&-0.25&-0.84&-0.39&-0.45&-1.01&-0.54&0.04&-0.16&-0.07&-2.31&-3.71&-5.30\\
    \cline{2-14}                         &KristenAndSara&-0.27&-0.32&0.12&-0.48&-0.36&-0.47&0.02&-0.43&0.07&-3.31&-1.79&-2.44\\
    \cline{1-14}
    \multicolumn{2}{|c|}{{AVERAGE}}&{-0.19}&{-0.26}&{-0.12}&{-0.45}&{-0.34}&{-0.27}&{-0.06}&{-0.15}&{-0.12}&{-2.28}&{-1.74}&{-2.18}\\
    \hline
   \end{tabular}
\end{center}
\end{table*}

\subsection{Coding Performance Comparison}
The experiments are conducted on the platform VTM 5.0 following the default AI configuration and the \textbf{Common Test Conditions} (\textbf{CTC}) \cite{CTC}. The workstation is equipped with the Intel Core i7-4790 CPU @2.60 GHz, Windows 7 Enterprise 64-bit operating systems for video coding. The original VTM 5.0 is regarded as the anchor for coding performance comparison, which is evaluated by \textbf{Bj${\phi}$ntegaard Delta Bit Rate} (\textbf{BD-BR}) \cite{BDBR}. Twenty-two sequences with various contents and resolutions, different from the training set, are utilized in the experiments.

Table \ref{table5} illustrates the values of BPM under the proposed method. Two sequences of each class are utilized for this experiment, which are identical to those in Table \ref{table1}. During intra coding, the total bits of intra modes and the number of intra blocks are collected for BPM calculation when the QP value equals to \{22, 27, 32, 37\}. Compared with those shown in Table \ref{table1}, the average values of BPM are changed from 3.35, 3.48, 3.44, and 3.39 to 2.38, 2.35, 2.22, and 2.20 under four QP settings, respectively. In general, the coding bit saving of intra mode can be calculated as follows regardless of residue and other information,
\begin{equation}
\eta' = \frac{\alpha - {\alpha}'}{\alpha} \times 100 \%,
\end{equation}
where $\alpha$ is the original value of BPM and $\alpha'$ is the current value of BPM. Accordingly, the bit saving of intra mode can reach 30.4\%, 33.2\%, 36.2\%, and 35.3\% on average under four QP settings, respectively.

Three state-of-the-art works are adopted for coding performance comparison. Narsallah's scheme \cite{8803773} derives the intra mode with gradient histogram and the one with the highest probability is determined eventually. Abdoli's scheme \cite{9102799} produces a new intra prediction result with weighted intra modes from the top-2 highest probability in gradient histogram. Li's scheme \cite{9115218} re-constructs the MPM list with short and long range correlations. These three works are optimized from different directions, and related to the proposed method, which can be compared in terms of coding performance. The comparison is illustrated in Table \ref{table6}. \par

For Narsallah's scheme \cite{8803773}, it reduces 0.19\%, 0.26\%, and 0.12\% bit rate on average for Y, U, and V components, respectively. 0.45\%, 0.34\%, and 0.27\% bit rates are saved for Y, U, and V components in the Abdoli's scheme \cite{9102799}. For Li's scheme \cite{9115218}, it achieves 0.06\%, 0.15\%, and 0.12\% bit rate reduction on average for luma and two chroma components, respectively. Regarding the proposed method, the bit rate reduction reaches 2.28\%, 1.74\%, and 2.18\% on average for luma and two chroma components, respectively. From this comparison, it can be observed that the proposed method is better than other three methods. Compared with Narsallah's scheme \cite{8803773}, the proposed method not only adopts the existing hand-crafted features, but also learns features in highly dimensional space for the intra mode derivation.

\begin{table*}[t]\caption{Performance evaluation in terms of BD-BR with different QP settings. (Unit: \%)} \label{table7}
\footnotesize
\begin{center}
    \begin{tabular}{|c|c|c|c|c|c|c|c|c|c|}
    \cline {1-8} \multirow{2}{*}{Class}&\multirow{2}{*}{Sequence}&\multicolumn{3}{c|}{Small QP \{11,16,21,26\}}&\multicolumn{3}{c|}{Large QP \{33,38,43,48\}}\\
    \cline {3-8} &&{Y}&{U}&{V}&{Y}&{U}&{V}\\
    \hline
    \hline
               \multirow{3}{*}{A1}      &Tango2&-0.76&-0.32&0.11&-3.68&-4.01&-3.88\\
    \cline{2-8}                         &FoodMarket4&-0.31&-0.77&-0.60&-3.24&-2.89&-3.28 \\
    \cline{2-8}                         &Campfire&-1.00&-0.66&-0.67&-4.37&-2.63&-3.27 \\
    \cline{1-8}\multirow{3}{*}{A2}      &CatRobot1&-0.79&-0.25&-0.42&-4.37&-2.63&-3.27 \\
    \cline{2-8}                         &DaylightRoad2&-0.44&-0.18&-0.03&-5.03&-4.61&-5.28 \\
    \cline{2-8}                         &ParkRunning3&-0.26&-0.26&-0.23&-2.33&-1.60&-1.72 \\
    \cline{1-8}\multirow{5}{*}{B}       &MarketPlace&-0.58&-0.57&-0.04&-2.97&-5.29&0.23 \\
    \cline{2-8}                         &RitualDance&-0.29&-0.58&-1.14&-4.87&-6.58&-4.68 \\
    \cline{2-8}                         &Cactus&-0.60&-0.49&-0.46&-4.15&-1.81&-3.41 \\
    \cline{2-8}                         &BasketballDrive&-0.60&0.01&-0.86&-3.65&-4.40&-4.16 \\
    \cline{2-8}                         &BQTerrace&-0.63&-0.36&-0.47&-3.70&-3.57&-6.81 \\
    \cline{1-8}\multirow{4}{*}{C}       &BasketballDrill&-0.87&-1.63&-1.20&-3.03&-5.43&0.93 \\
    \cline{2-8}                         &BQMall&-1.00&-0.59&-0.86&-4.16&-4.52&-5.75 \\
    \cline{2-8}                         &PartyScene&-0.84&-0.51&-0.58&-3.71&-0.98&-6.77 \\
    \cline{2-8}                         &RaceHorsesC&-0.70&-0.55&-0.57&-3.35&-2.35&-4.11 \\
    \cline{1-8}\multirow{4}{*}{D}       &BasketballPass&-0.66&0.53&-1.87&-1.91&-6.51&-3.23 \\
    \cline{2-8}                         &BQSquare&-0.89&-0.79&-1.62&-3.84&-10.4&-9.82 \\
    \cline{2-8}                         &BlowingBubbles&-0.76&-1.34&-0.59&-2.88&-0.62&1.48 \\
    \cline{2-8}                         &RaceHorses&-0.89&-0.79&-1.62&-3.84&-10.4&-9.82 \\
        \cline{1-8}\multirow{3}{*}{E}   &FourPeople&-0.87&-0.71&-0.97&-2.96&-2.23&-4.26 \\
    \cline{2-8}                         &Johnny&-1.21&-1.08&-0.89&-4.20&-3.56&-4.80 \\
    \cline{2-8}                         &KristenAndSara&-0.62&-0.64&-1.15&-4.36&-3.06&-2.66 \\
    \cline{1-8}
    \multicolumn{2}{|c|}{{AVERAGE}}&{-0.71}&{-0.57}&{-0.76}&{-3.64}&{-4.15}&{-4.00}\\
    \hline
   \end{tabular}
\end{center}
\end{table*}

In addition, the test sequences are encoded under the small QP setting \{11, 16, 21, 26\} and large QP setting \{33, 38, 43, 48\} to evaluate the performance of the proposed method. It should be noted that the neural network is not re-trained. The coding performance is shown in Table \ref{table7}. The bit rate reductions can reach 0.71\% and 3.64\% for luma component under the small and large QP settings, respectively. Compared with the results in Table \ref{table6}, the performance of normal QP setting is a little worse than that of large QP setting and better than that of small QP setting. The reason is that the percentage of coding bits of intra mode in a frame becomes large as QP value increases, and vice versa. Consequently, in the low bit rate scenario, the compression efficiency has been greatly improved for the proposed method.

\begin{figure*}[t]
  \centering
 \subfigure[BasketballPass]{
    \label{fig9:subfig:a} 
    \includegraphics[width=0.32\textwidth, height=0.21\textwidth]{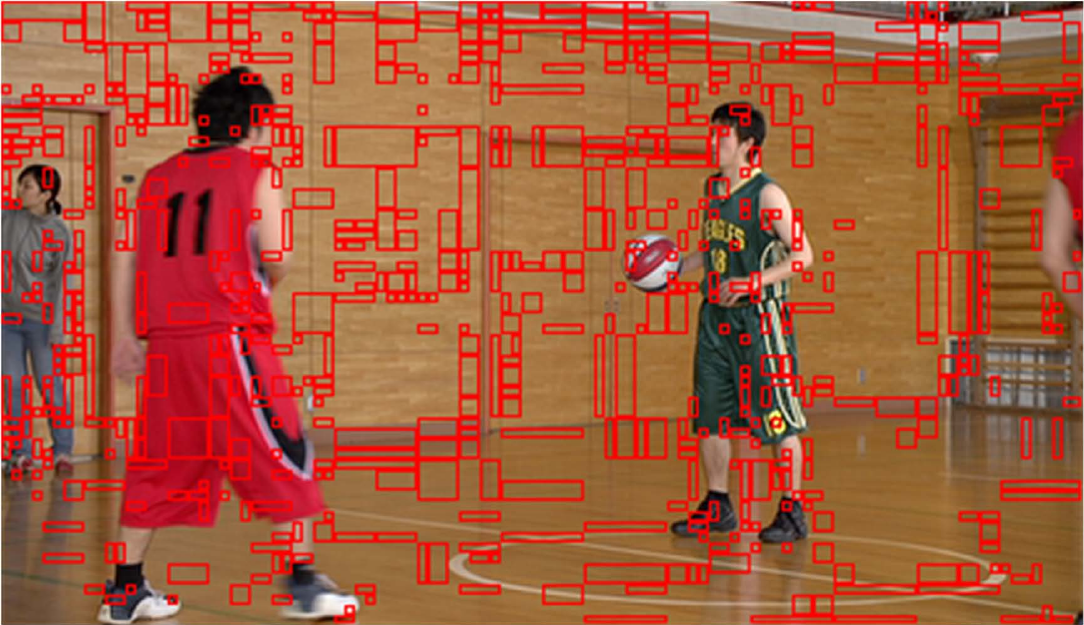}}
 \subfigure[BQSquare]{
    \label{fig9:subfig:d} 
    \includegraphics[width=0.32\textwidth, height=0.21\textwidth]{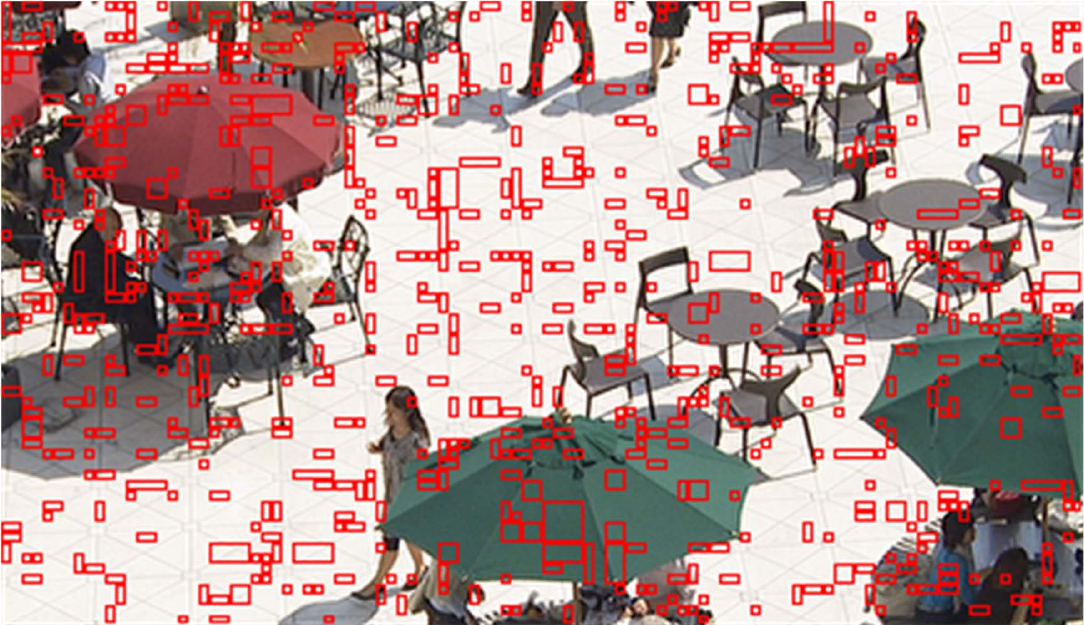}}
  \subfigure[BQMall]{
    \label{fig9:subfig:c} 
    \includegraphics[width=0.32\textwidth, height=0.21\textwidth]{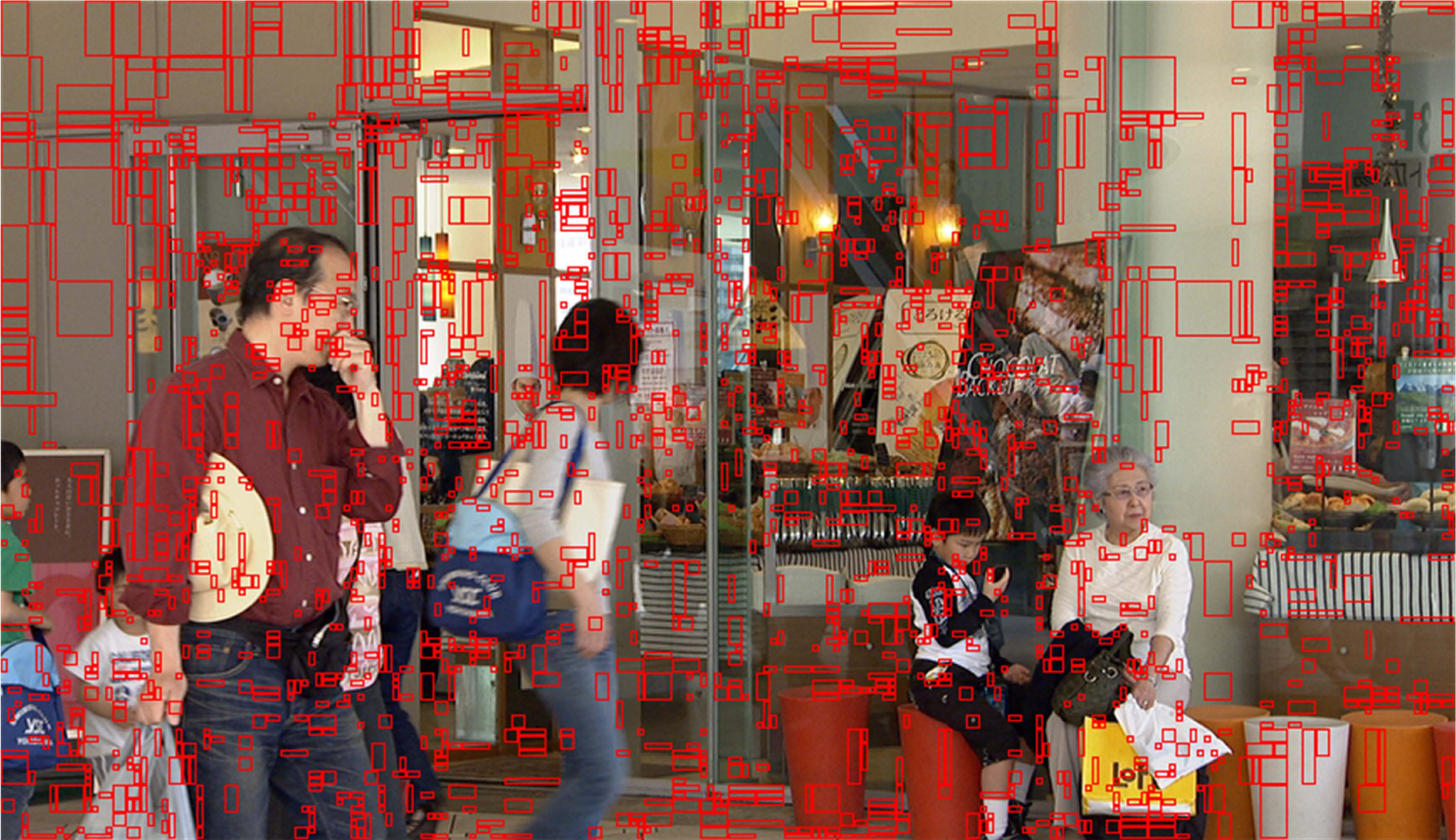}}
  \subfigure[BasketballDrill]{
    \label{fig9:subfig:d} 
    \includegraphics[width=0.32\textwidth, height=0.21\textwidth]{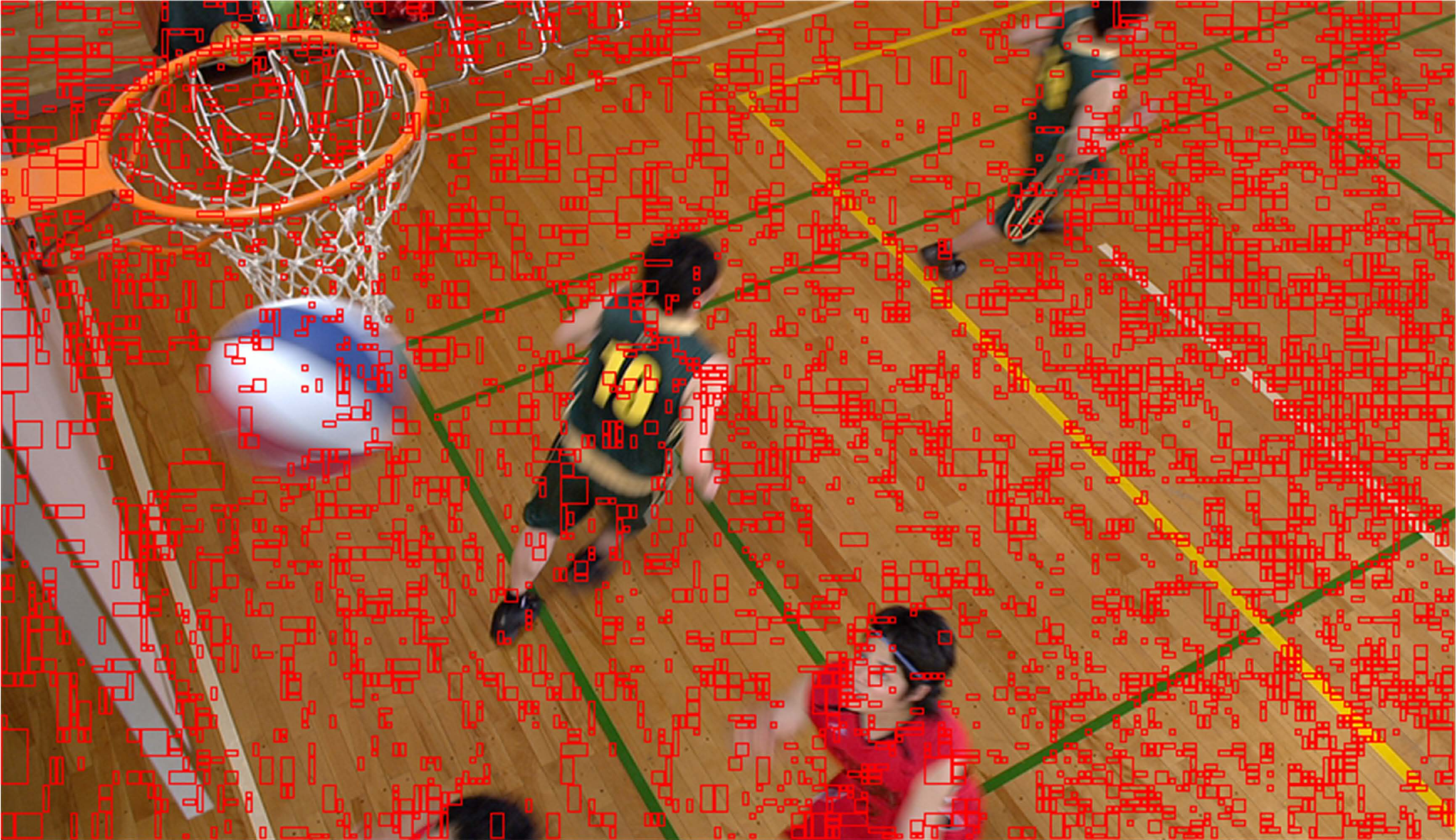}}
  \subfigure[FourPeople]{
    \label{fig9:subfig:e} 
    \includegraphics[width=0.32\textwidth, height=0.21\textwidth]{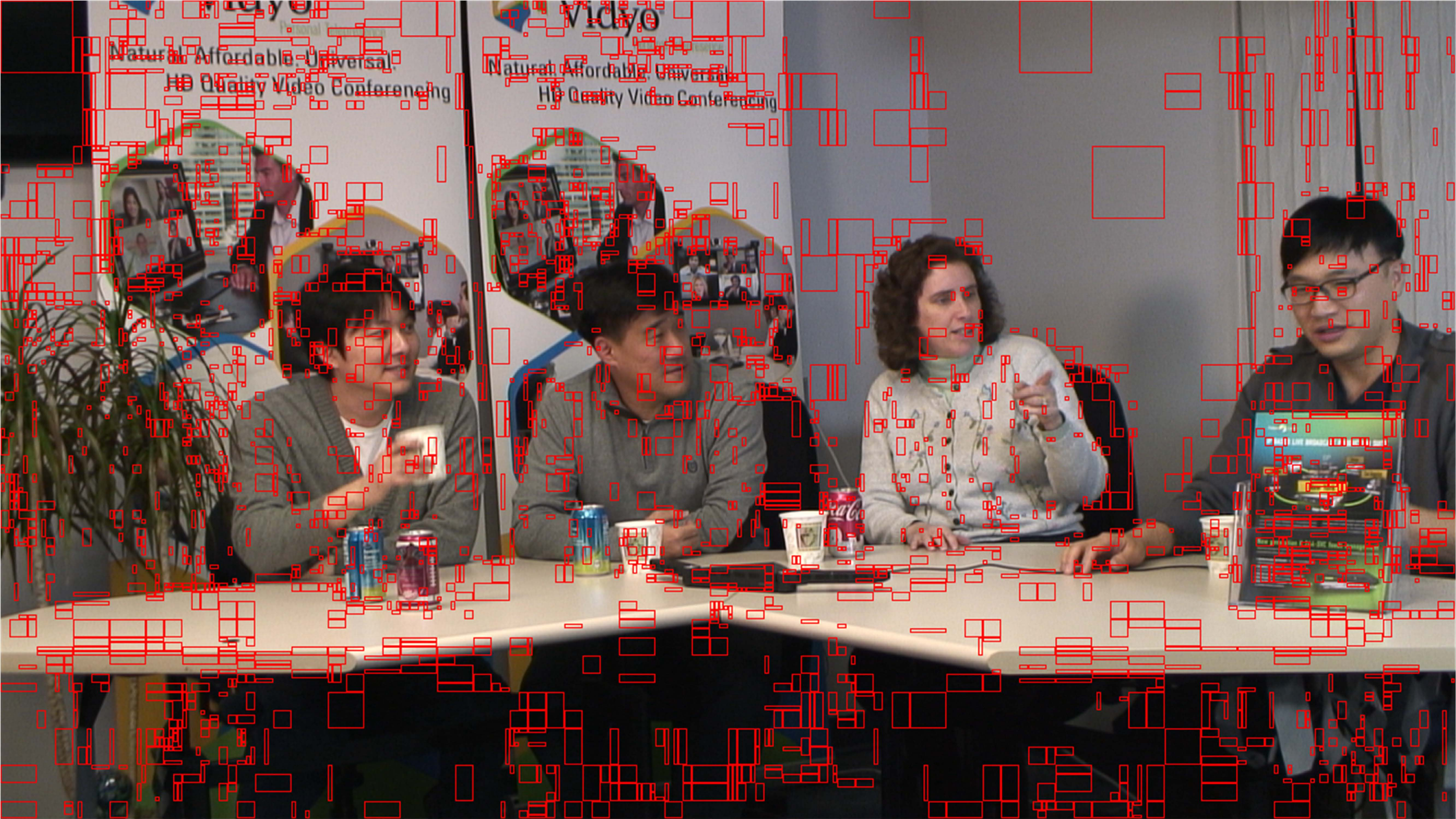}}
  \subfigure[Johnny]{
    \label{fig9:subfig:f} 
    \includegraphics[width=0.32\textwidth, height=0.21\textwidth]{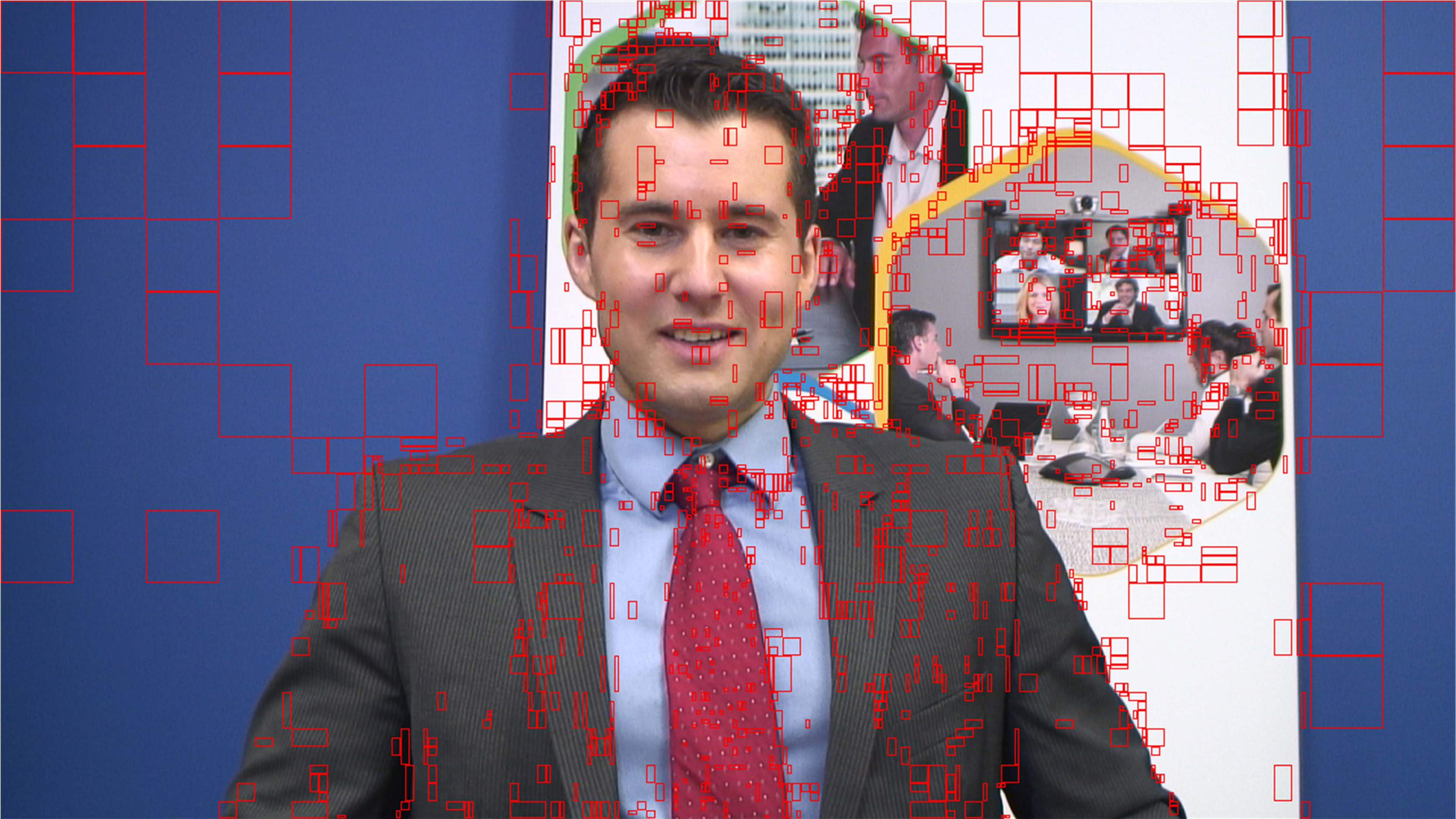}}
  \caption{DLIMD selected in the a frame. (They are resized to the same resolution for visualization.)}
  \label{Fig11} 
\end{figure*}

\begin{table}[t]\caption{Percentage of the proposed method selection. (Unit: \%)} \label{table8}
\footnotesize
\begin{center}
    \begin{tabular}{|c|c|c|c|c|c|}
    \cline {1-6} {Class}&{Sequence}&{QP=22}&{QP=27}&{QP=32}&{QP=37}\\
    \hline
    \hline
               \multirow{3}{*}{A1}      &Tango2&39.9&41.3&44.3&46.5\\
    \cline{2-6}                         &FoodMarket4&38.5&37.6&41.4&44.2\\
    \cline{2-6}                         &Campfire&48.0&48.6&49.8&53.5\\
    \cline{1-6}\multirow{3}{*}{A2}      &CatRobot1&41.3&47.3&50.5&50.6\\
    \cline{2-6}                         &DaylightRoad2&45.4&50.9&52.8&52.4\\
    \cline{2-6}                         &ParkRunning3&43.6&46.5&47.2&47.9\\
    \cline{1-6}\multirow{5}{*}{B}       &MarketPlace&41.3&44.5&47.1&48.9\\
    \cline{2-6}                         &RitualDance&43.8&45.9&50.2&52.2\\
    \cline{2-6}                         &Cactus&40.8&45.1&48.1&50.7\\
    \cline{2-6}                         &BasketballDrive&43.4&46.1&50.6&52.6\\
    \cline{2-6}                         &BQTerrace&41.7&48.4&51.7&54.4\\
    \cline{1-6}\multirow{4}{*}{C}       &BasketballDrill&57.8&60.3&59.3&56.1\\
    \cline{2-6}                         &BQMall&43.3&44.9&48.5&50.5\\
    \cline{2-6}                         &PartyScene&41.8&45.5&47.8&52.5\\
    \cline{2-6}                         &RaceHorsesC&39.8&43.5&45.4&49.9\\
    \cline{1-6}\multirow{4}{*}{D}       &BasketballPass&45.5&40.8&48.8&51.3\\
    \cline{2-6}                         &BQSquare&39.6&40.6&45.5&49.8\\
    \cline{2-6}                         &BlowingBubbles&41.5&42.8&47.7&48.7\\
    \cline{2-6}                         &RaceHorses&38.8&39.7&45.6&50.3\\
    \cline{1-6}\multirow{3}{*}{E}       &FourPeople&43.2&45.5&47.5&48.2\\
    \cline{2-6}                         &Johnny&42.3&45.0&49.2&50.0\\
    \cline{2-6}                         &KristenAndSara&42.2&43.3&47.4&49.4\\
    \cline{1-6}
    \multicolumn{2}{|c|}{{AVERAGE}}&{42.9}&{45.2}&{48.5}&{50.5}\\
    \hline
   \end{tabular}
\end{center}
\end{table}

As shown in Fig. \ref{Fig11}, the first frames of six sequences, including BasketballPass ($416\times240$), BQSquare ($416\times240$), BQMall ($832\times480$), BasketballDrill ($832\times480$), FourPeople ($1280\times720$), and Johnny ($1280\times720$), 
are used to demonstrate the proposed DLIMD selection in a frame, where the QP value is 22 and the coding blocks are marked as red color in case that the DLIMD is selected. It can be clearly observed that lots of coding blocks select the proposed DLIMD. Moreover, the quantitative results are presented in Table \ref{table8} under different QP settings and different sequences. The percentage of DLIMD selection is calculated by the ratio of selected area against the whole frame, which is represented by,
\begin{equation}
\Omega = \frac{\sum_{i=1}^{N}C_i \times w_i \times h_i}{\sum_{i=1}^{N}w_i \times h_i} \times 100\%,
\end{equation}
where $N$ indicates the number of coding blocks in a frame, $C_i$ indicates the DLIMD selection, $C_i = 0$ if the current coding block does not select DLIMD, $w_i$ and $h_i$ are the width and height of the current coding block.
From this table, the percentage can reach 42.9\%, 45.2\%, 48.5\%, and 50.5\% on average under four QP settings, respectively. It indicates that the coding performance can be efficiently improved. \par

\subsection{Influence of Learned and Hand-crafted Features}
The individual influence of learned features and hand-crafted features in the proposed architecture (shown in Fig. \ref{Fig6}) is analyzed. Four cases are presented, i.e., (1) H: the module of learning features is removed, only the hand-crafted features are used for intra mode derivation; (2) L: the hand-crafted features are removed, only the learned features are used for intra mode derivation; (3) H'+L: both the hand-crafted features (excluding gradient histogram) and learned features are used for intra mode derivation; (4) H+L: both the hand-crafted features and learned features are used for intra mode derivation.  \par

\begin{figure*}[t]
  \centering
 \subfigure[Training loss]{
    \label{fig9:subfig:a} 
    \includegraphics[width=0.38\textwidth]{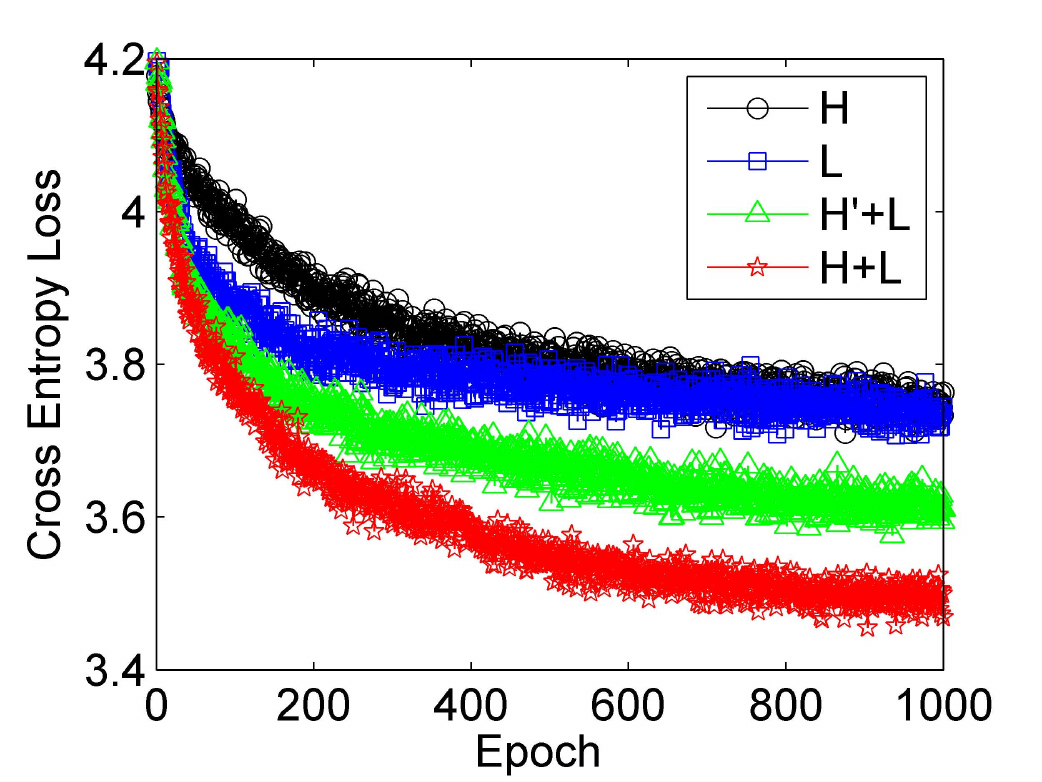}}
 \subfigure[Validation classification accuracy]{
    \label{fig9:subfig:d} 
    \includegraphics[width=0.38\textwidth]{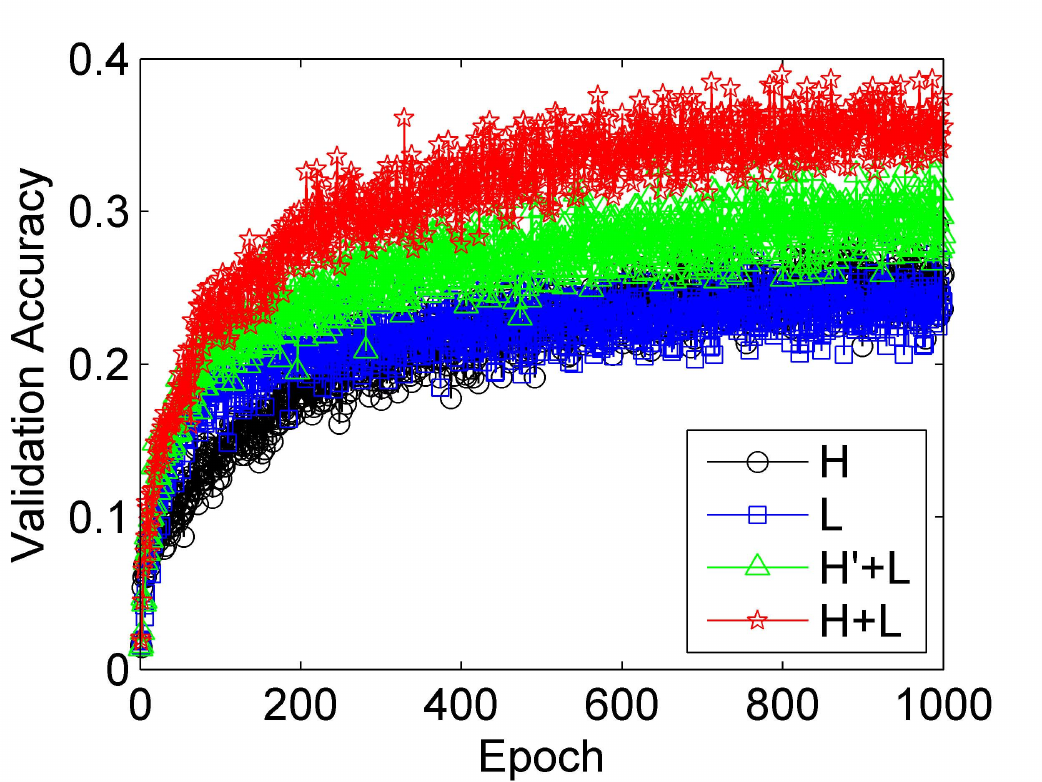}}
  \caption{Comparison of four cases with hand-crafted and learned features. }
  \label{Fig8} 
\end{figure*}

With the same training samples claimed in Section 3.3, three more neural networks are trained separately according to the listed cases. Fig. \ref{Fig8} illustrates the comparison of these four cases in terms of training loss and validation classification accuracy. The classification accuracy is calculated as follows,
\begin{equation}
P = \frac{1}{N}\sum_{i=1}^N \delta_i \times 100\%,
\end{equation}
where $N$ is the number of testing samples, $\delta_i=1$ in case that the difference between predicted label and ground truth is less than a pre-defined threshold $\Delta$, i.e., $\lVert \texttt{argmax}(\textbf{y}^i) - \texttt{argmax}(\textbf{O}^f_5)\rVert \leq \Delta$, otherwise $\delta_i=0$. \texttt{argmax}() returns the position of maximum value in a vector, $\textbf{y}^i$ is the ground truth represented in the one-hot manner, $\textbf{O}^f_5$ is the output of intra mode derivation network. Here, the value of $\Delta$ is set as 0. The cases of separate hand-crafted and learned features both converge at round 3.8 cross entropy loss and achieve about 25\% validation classification accuracy, the case of H'+L converges at about 3.6 cross entropy loss and achieves about 30\% validation classification accuracy, while the combination of hand-crafted and learned features converges at 3.5 cross entropy loss and achieves about 35\% validation classification accuracy. \par

From these results, it can be obviously observed that the hand-crafted and learned combination case achieves the best performance when compared with other three. The reasons are that although CNN is able to extract high-level features and latent representation, the hand-crafted features still can provide useful information and compensate the limitation of learned features. For example, the intra modes of neighbors from spatial domain, which cannot be learned from the feature learning network, play an important role for intra mode derivation.

\subsection{Ablation Study of Architecture}
\begin{figure}[t]
  \centering
 \subfigure[Feature learning network.]{
    \label{fig9:subfig:a} 
    \includegraphics[width=0.36\textwidth]{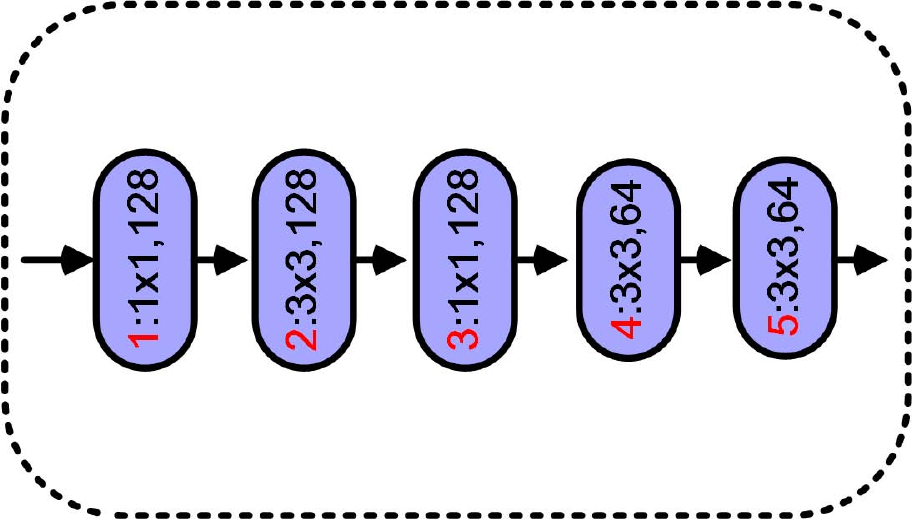}}
 \subfigure[Intra mode derivation network.]{
    \label{fig9:subfig:d} 
    \includegraphics[width=0.36\textwidth]{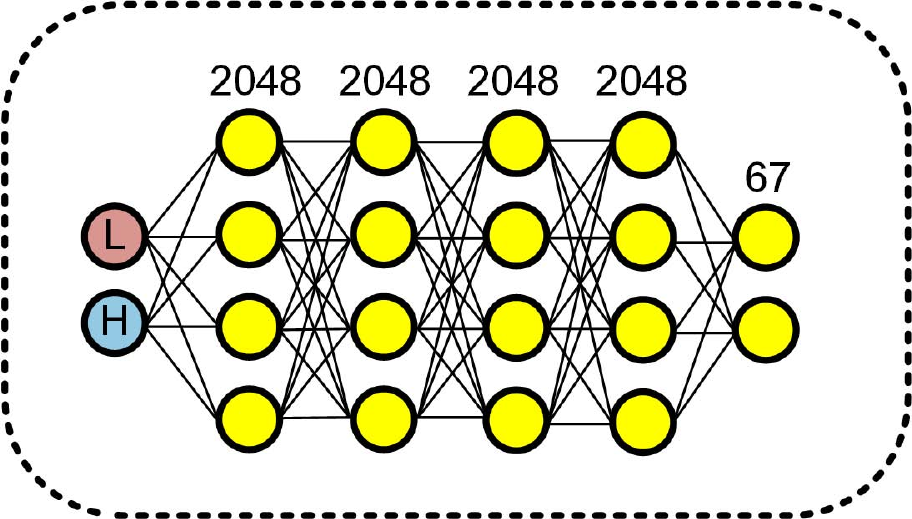}}
  \caption{Alternative network. (H: hand-crafted feature, L: learned feature)}
  \label{Fig9} 
\end{figure}

In addition, we aim to further analyze the impact of modules in the network architecture. Alternative networks are designed, and illustrated in Fig. \ref{Fig9}. Different from the proposed network shown in Fig. \ref{Fig6}, the convolutional layers are placed in the serial manner, the number of feature maps in the first three layers is 128 which matches the input of convolutional layers 2a, 2b, 3a, 3b in Fig. \ref{Fig6}, the kernel sizes of the first and third convolutional layers are set as $1\times1$ and the others are $3\times3$, the hand-crafted features are only combined to the first fully connected layer.\par

Three configurations are listed for comparison, i.e., Case A: feature learning network in Fig. \ref{Fig6} and intra mode derivation network in Fig. \ref{Fig6}; Case B: feature learning network in Fig. \ref{Fig6} and intra mode derivation network in Fig. \ref{Fig9}; Case C: feature learning network in Fig. \ref{Fig9} and intra mode derivation network in Fig. \ref{Fig6}. It should be noted that Case A is the proposed one. Here, two more networks for Cases B and C are trained with the same samples. The results are compared in terms of multi-class classification accuracy. Two test sequences from each class defined in the CTC \cite{CTC} are employed, i.e., BasketballPass ($416\times240$), BlowingBubbles ($416\times240$), BQMall ($832\times480$), BasketballDrill ($832\times480$), FourPeople ($1280\times720$), Johnny ($1280\times720$), BasketballDrive ($1920\times1080$), BQTerrace ($1920\times1080$), Tango2 ($3840\times2160$) and FoodMarket4 ($3840\times2160$). For each sequence, $800\times67\times4$ samples are selected under four QP settings. Table \ref{table4} illustrates the experimental results, and the classification accuracy is calculated by Eq. (10). In Table \ref{table4}, under this condition of $\Delta = 0$, the multi-class classification accuracies are 34.8\%, 31.4\%, and 32.6\% on average for Cases A, B, and C, respectively. As such, we can conclude that the hand-crafted features combined to every fully connected layer and different kernel sizes placed in the parallel manner in convolutional layers can achieve better performance. \par

\begin{table*}[t]\caption{Multi-class classification accuracy. (Unit: \%)} \label{table4}
\footnotesize
\begin{center}
    \begin{tabular}{|c|c|c|c|c|c|c|c|c|c|c|c|c|c|}
    \cline {1-14} \multirow{2}{*}{Class}&\multirow{2}{*}{Sequence}&\multicolumn{4}{c|}{Case A (proposed)}&\multicolumn{4}{c|}{Case B}&\multicolumn{4}{c|}{Case C}\\
    \cline {3-14} &&{$\Delta$=0}&{$\Delta$=1}&{$\Delta$=3}&{$\Delta$=5}&{$\Delta$=0}&{$\Delta$=1}&{$\Delta$=3}&{$\Delta$=5}&{$\Delta$=0}&{$\Delta$=1}&{$\Delta$=3}&{$\Delta$=5}\\
    \hline
    \hline
                \multirow{2}{*}{A}   &Tango2&31.8&45.3&59.8&68.0&28.4&41.6&56.7&65.3&29.6&43.3&58.1&67.1 \\
    \cline{2-14}                     &FoodMarket4&34.5&50.9&67.0&75.3&31.1&47.8&64.2&73.3&31.8&49.1&65.9&74.4\\
    \cline{1-14}\multirow{2}{*}{B}   &BasketballDrive&37.2&52.6&65.6&72.1&34.0&49.6&62.5&69.5&34.5&49.3&63.1&70.2\\
    \cline{2-14}                     &BQTerrace&32.8&47.3&58.6&64.9&28.7&43.0&54.4&61.3&31.0&46.1&58.0&64.7 \\
    \cline{1-14}\multirow{2}{*}{C}   &BQMall&35.6&50.9&62.3&67.5&32.9&47.4&58.2&64.1&33.9&50.0&60.9&66.8 \\
    \cline{2-14}                     &BasketballDrill&37.3&55.0&66.9&72.4&33.8&52.4&64.6&70.0&35.4&54.0&66.9&72.5 \\
    \cline{1-14}\multirow{2}{*}{D}   &BlowingBubbles&32.1&47.2&61.1&67.8&29.2&44.1&58.2&65.7&30.2&46.0&60.8&68.3 \\
    \cline{2-14}                     &BasketballPass&34.7&50.8&63.4&69.3&32.0&48.0&59.5&65.8&32.5&49.9&62.2&68.2 \\
    \cline{1-14}\multirow{2}{*}{E}   &FourPeople&34.1&49.3&61.5&67.7&29.8&44.8&56.8&62.9&32.2&47.5&59.7&66.1 \\
    \cline{2-14}                     &Johnny&37.6&54.7&68.4&75.1&34.2&51.5&65.5&72.7&35.2&52.5&66.6&73.4 \\
    \cline{1-14}
    \multicolumn{2}{|c|}{{AVERAGE}}&{34.8}&{50.4}&{63.5}&{70.0}&{31.4}&{47.0}&{60.1}&{67.1}&{32.6}&{48.7}&{62.2}&{69.1}\\
    \hline
   \end{tabular}
\end{center}
\end{table*}

\begin{figure*}[t]
  \centering
 \subfigure[BasketballDrive]{
    \label{fig9:subfig:d} 
    \includegraphics[width=0.3\textwidth]{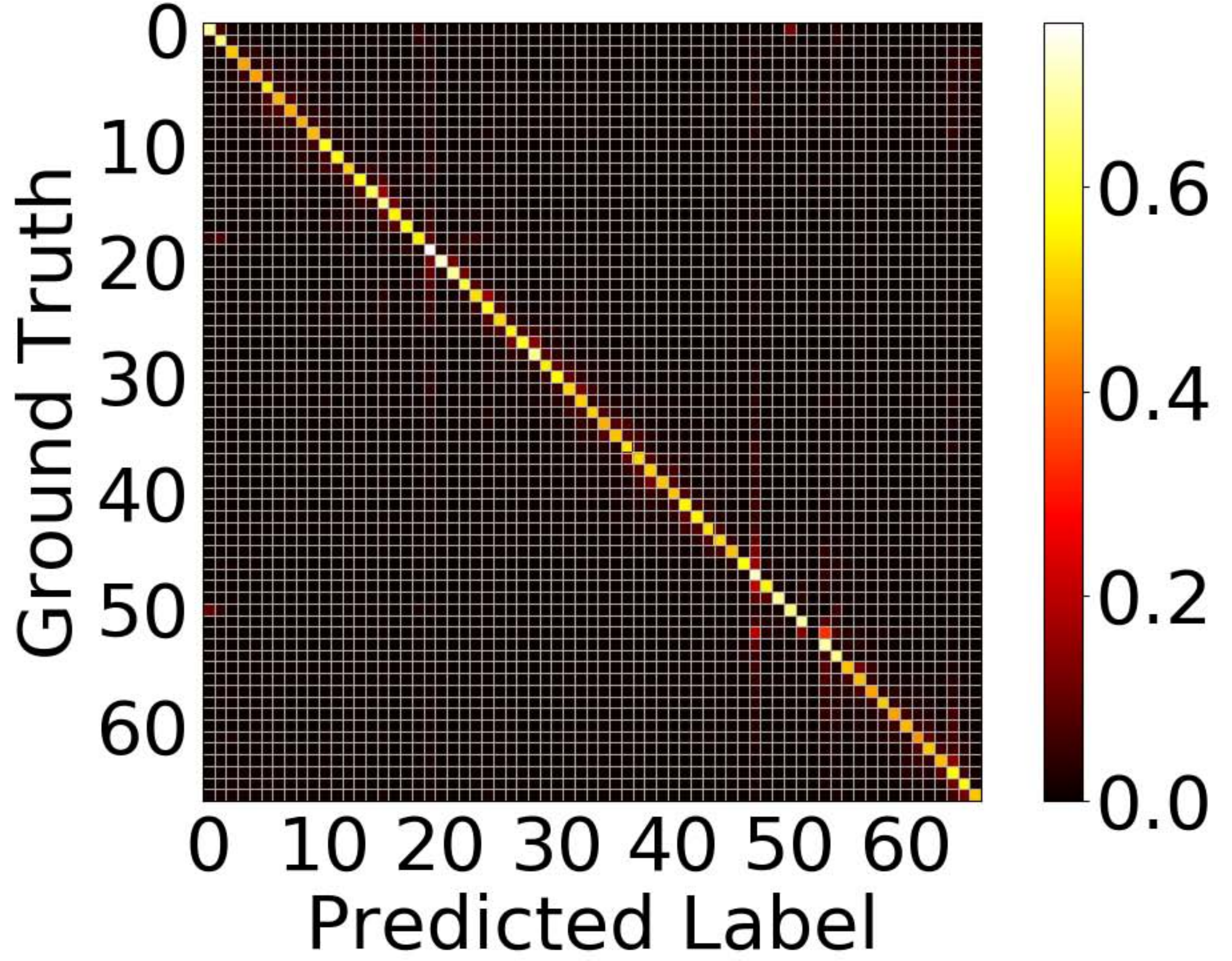}}
  \subfigure[BQMall]{
    \label{fig9:subfig:c} 
    \includegraphics[width=0.3\textwidth]{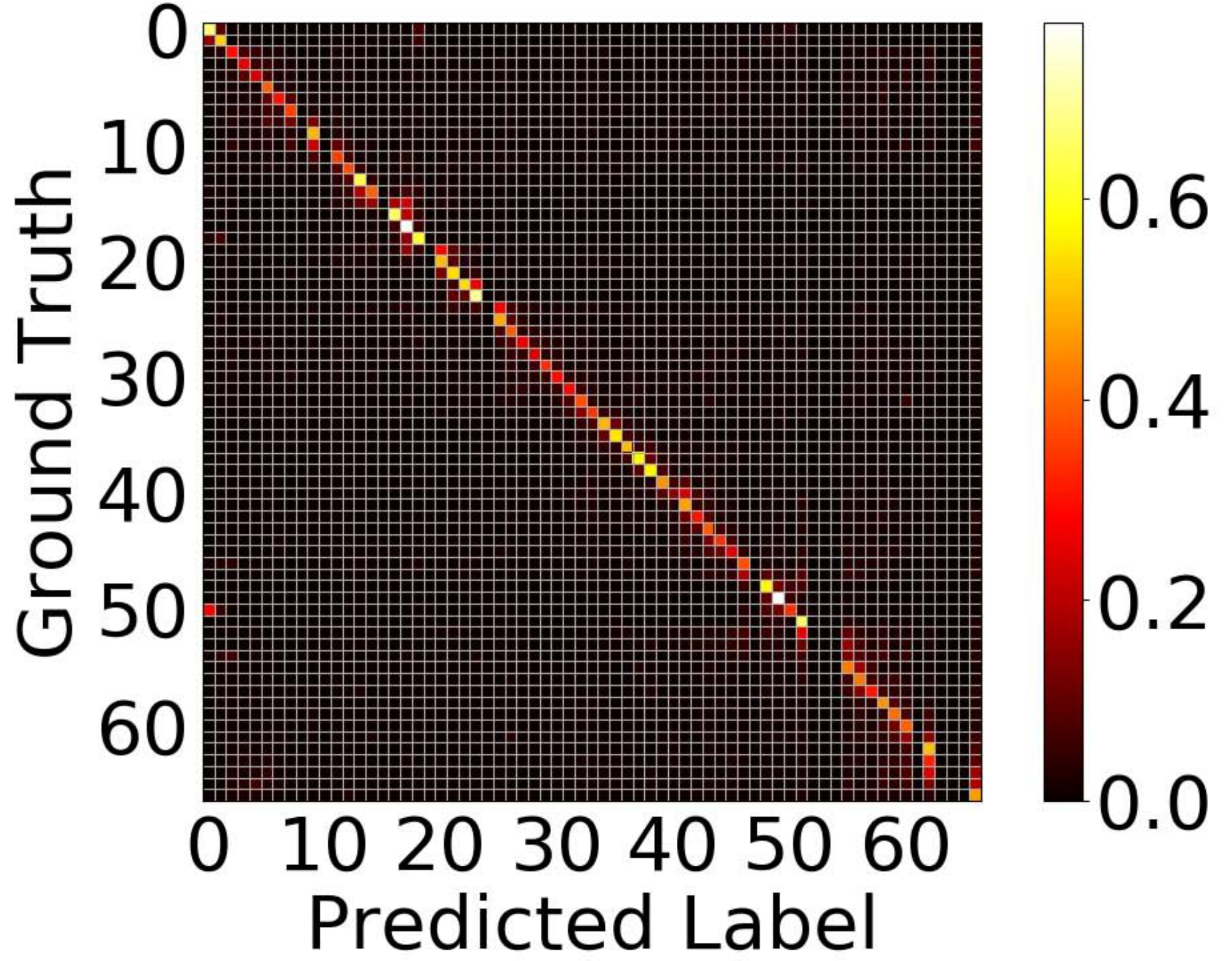}} \\
  \subfigure[BlowingBubbles]{
    \label{fig9:subfig:d} 
    \includegraphics[width=0.3\textwidth]{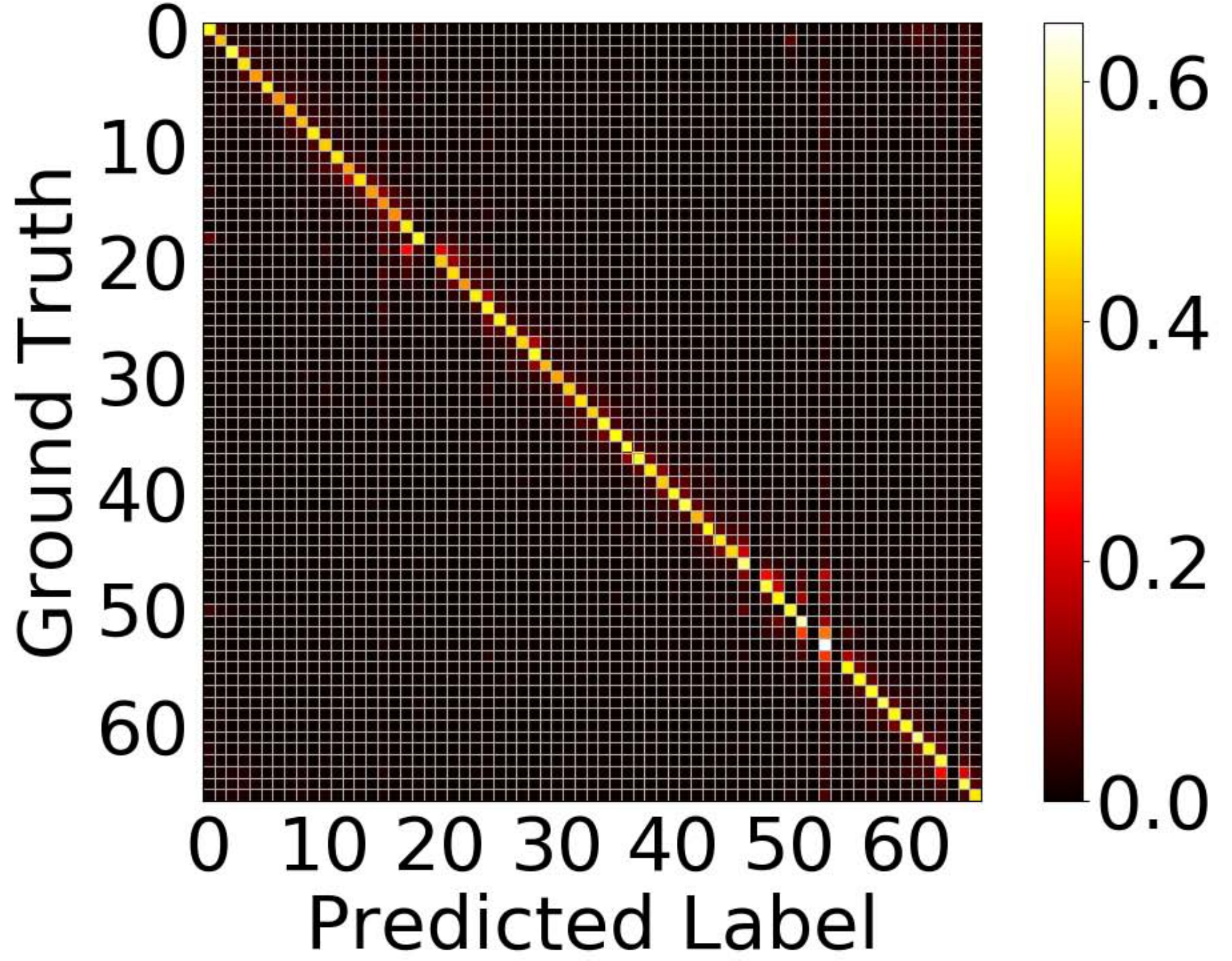}}
  \subfigure[FourPeople]{
    \label{fig9:subfig:d} 
    \includegraphics[width=0.3\textwidth]{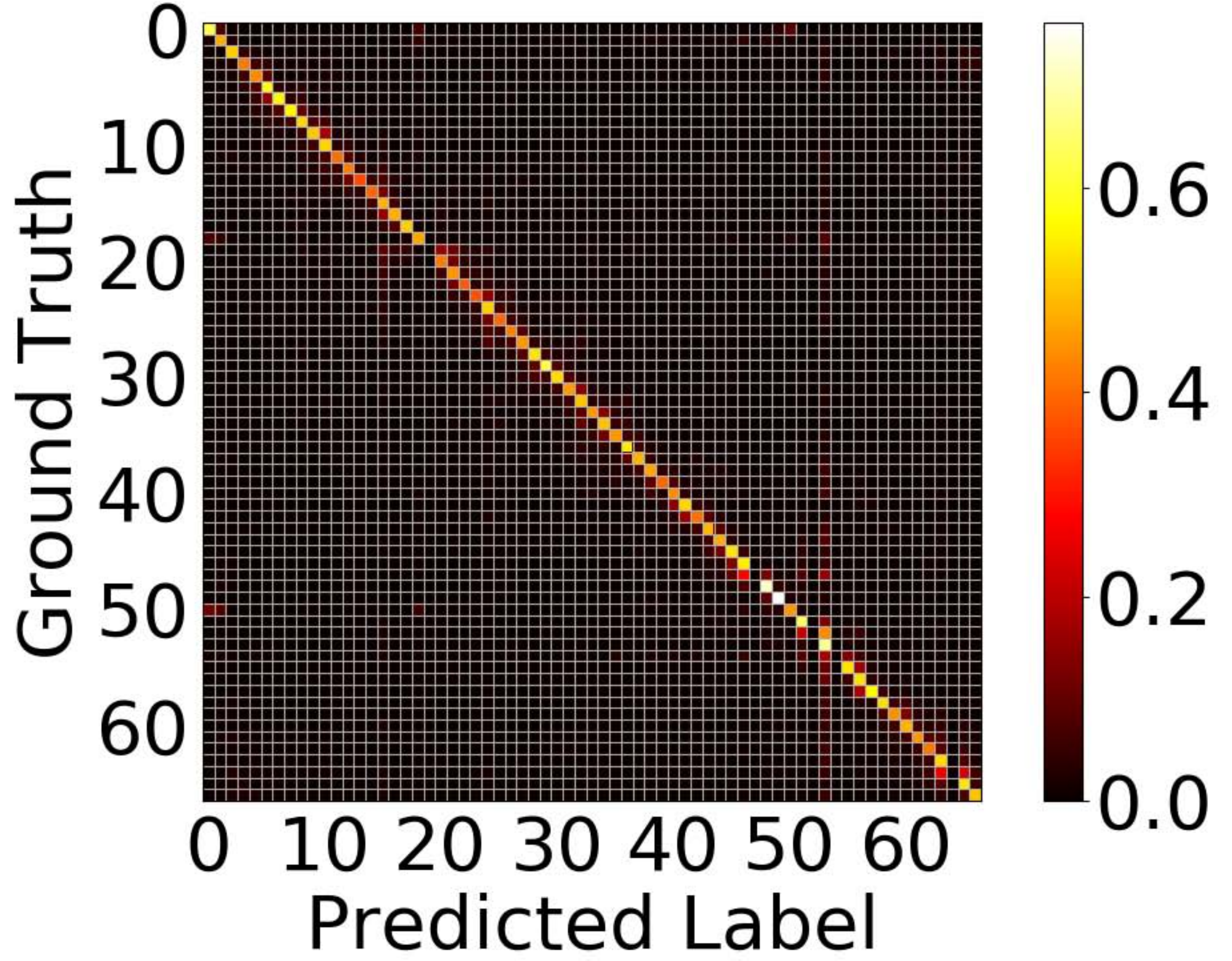}}
  \caption{Confusion matrix of multi-class classification under Case A.}
  \label{Fig10} 
\end{figure*}

In addition, the normalized confusion matrices of Case A are illustrated in Fig.\ref{Fig10}. The horizontal is predicted label and the vertical is ground truth. It can be observed that the difference between ground truth and predicted label is limited. For the proposed one (Case A), the average classification accuracies under four conditions are 34.8\%, 50.4\%, 63.5\%, and 70.0\% in Table \ref{table4}, respectively. Although there are some differences between predicted label and ground truth under the conditions of $\Delta = \{1, 3, 5\}$, the intra prediction results may be similar, 
and the RDO will be performed to balance the distortion and coding bits during video coding. Therefore, the coding gains can still be achieved with limited difference between predicted label and ground truth.

\subsection{Computational Complexity Analyses}
Additionally, the coding/decoding time of video codec equipped with the DLIMD is compared with that of the anchor, which is calculated by,
\begin{equation}
\Delta T_m = \frac{1}{4}\sum_{i=1}^{4}{\frac{T_{\Psi}^m(QP_i)}{T_{c}^m(QP_i)}},
\end{equation}
where $T_{c}^m(QP_i)$ is the coding/decoding time of the anchor under $QP_i$, and $T_{\Psi}^m(QP_i)$ is the coding/decoding time of the video codec equipped with proposed method under $QP_i$, $m \in$ \{coding, decoding\}. Compared with the anchor, the values of computational complexity of the proposed method are 33.6 times, 140.3 times under CPU+GPU platform and 231.3 times, 604.8 times under CPU platform on average for video coding and decoding, respectively. The computational complexity is a great challenge. In the video codec, the DLIMD is performed in the variable coding blocks and the convolutional/fully connected operations in the neural network result in high complexity.

For other deep learning based schemes \cite{9266121}\cite{8794555} that focus on the optimization of intra prediction, the values of computational complexity are 9.87 times, 87.4 times at the encoder side and 151.7 times, 124.5 times at the decoder side with respect to the anchor. The former and latter schemes with 1.92\% and 3.4\% bit rate reductions for luma component are performed on the platform of CPU and CPU+GPU, respectively. For the conventional schemes \cite{8803773}\cite{9102799}\cite{9115218} whose compression efficiencies have been compared in Table \ref{table6}, the values of encoding and decoding complexity are 109\%, 111\%, 101\%  and 104\%, 105\%, 100\% with respect to the anchor. It can be found that the computational complexity of deep learning based schemes including the proposed one is much higher than that of conventional schemes.

\begin{table*}[t]\caption{Trade-off between computational complexity and compression efficiency on the platform of CPU+GPU. } \label{table9}
\footnotesize
\begin{center}
    \begin{tabular}{|C{0.48cm}|C{1.68cm}|C{0.7cm}|C{0.7cm}|C{0.7cm}|C{0.7cm}|C{0.7cm}|C{0.7cm}|C{0.7cm}|C{0.7cm}|C{0.7cm}|C{0.7cm}|}
    \cline {1-12} \multirow{3}{*}{Class}&\multirow{3}{*}{Sequence}&\multicolumn{5}{c|}{DLIMD}&\multicolumn{5}{c|}{DLIMD-L}\\
    \cline {3-12} &&\multicolumn{3}{c|}{BDBR (\%)}&\multicolumn{2}{c|}{Complexity}&\multicolumn{3}{c|}{BDBR (\%)}&\multicolumn{2}{c|}{Complexity}\\
    \cline {3-12} &&{Y}&{U}&{V}&{Encode}&{Decode}&{Y}&{U}&{V}&{Encode}&{Decode}\\
    \hline
    \hline
                \multirow{2}{*}{A}   &Tango2&-2.50&-2.77&-2.20&36.8&144.7&-1.71&-2.53&-1.30&28.8&38.1 \\
    \cline{2-12}                     &FoodMarket4&-2.65&-1.66&-1.14&25.8&134.9&-1.92&-1.08&-0.39&19.9&39.1 \\
    \cline{1-12}\multirow{2}{*}{B}   &BasketballDrive&-2.42&-2.68&-2.04&36.6&135.3&-1.55&-1.70&-2.27&29.0&35.3 \\
    \cline{2-12}                     &BQTerrace&-1.88&-1.77&-2.40&37.0&167.9&-0.73&-0.97&-0.33&30.3&36.2 \\
    \cline{1-12}\multirow{2}{*}{C}   &BQMall&-2.89&-1.93&-1.36&33.9&138.2&-2.11&-0.36&-1.14&27.2&56.3 \\
    \cline{2-12}                     &BasketballDrill&-2.29&1.18&-2.62&33.4&176.9&1.40&0.97&-0.26&27.4&37.1 \\
    \cline{1-12}\multirow{2}{*}{D}   &BlowingBubbles&-2.09&-1.36&-1.67&30.4&136.8&-1.32&-0.35&-1.77&25.9&38.6 \\
    \cline{2-12}                     &BasketballPass&-1.67&-2.31&-4.17&32.2&155.8&-0.46&1.77&-1.65&26.7&43.0 \\
   \cline{1-12}\multirow{2}{*}{E}   &FourPeople&-3.21&-2.61&-2.28&42.9&136.5&-2.44&-2.81&-1.54&34.1&46.1 \\
    \cline{2-12}                     &Johnny&-2.31&-3.71&-5.30&38.1&131.2&-0.81&-0.73&-3.52&30.4&38.2 \\
    \cline{1-12}
    \multicolumn{2}{|c|}{{AVERAGE}}&{-2.39}&{-1.96}&{-2.52}&{34.7}&{145.8}&{-1.17}&{-0.78}&{-1.42}&{27.9}&{40.8}\\
    \hline
   \end{tabular}
\end{center}
\end{table*}

Generally, to accelerate deep learning based schemes, the strategies include SIMD optimization, neural network quantization, and parameters/layers pruning. The first two strategies require the support/optimization from hardware devices. Therefore, the third one is adopted to investigate the trade-off between computational complexity and compression efficiency. One more architecture (denoted as DLIMD-L) is designed by reducing the parameters, i.e., the output of last layer in feature learning network is changed from 64 to 16, and the number of nodes of hidden layers except the last one in intra mode derivation network is changed from 2048 to 128. According to the definition of FLOPs \cite{MolchanovTKAK17}, the computational complexity can be largely reduced. DLIMD-L is trained with the same training set as DLIMD. The coding experiments are performed on the platform of CPU+GPU and the results are presented in the Table \ref{table9}. It can be observed that the coding efficiency of DLIMD-L is -1.17\% on average for luma component in terms of BD-BR, which is worse than that of DLIMD. The values of encoding and decoding complexity of DLIMD-L are 27.9 times and 40.8 times with respect to the anchor (VTM 5.0), where 6.8 times of encoding complexity and 105.0 times of decoding complexity are reduced. Although the computational complexity is still high, we believe that it can be optimized in the future.

\begin{table*}[t]\caption{Coding performance in terms of BD-BR on the latest platform of VTM 16.0 under AI, LDP, RA configurations. (Unit: \%)} \label{table10}
\footnotesize
\begin{center}
    \begin{tabular}{|C{0.48cm}|C{1.68cm}|C{0.55cm}|C{0.55cm}|C{0.55cm}|C{0.55cm}|C{0.55cm}|C{0.55cm}|C{0.55cm}|C{0.55cm}|C{0.55cm}|}
    \cline {1-11} \multirow{2}{*}{Class}&\multirow{2}{*}{Sequence}&\multicolumn{3}{c|}{AI Configuration}&\multicolumn{3}{c|}{LDP Configuration}&\multicolumn{3}{c|}{RA Configuration}\\
    \cline {3-11} &&{Y}&{U}&{V}&{Y}&{U}&{V}&{Y}&{U}&{V}\\
    \hline
    \hline
                \multirow{3}{*}{A1}      &Tango2&-2.48&-0.33&-1.60&-0.83&-0.81&-0.46&-1.18&-0.96&-0.10 \\
    \cline{2-11}                         &FoodMarket4&-1.87&-2.17&-1.36&-0.63&-1.43&0.05&-0.94&-0.38&-1.03 \\
    \cline{2-11}                         &Campfire&-2.37&-1.49&-2.72&-1.18&-0.95&-0.71&-1.82&-1.47&-1.32 \\
    \cline{1-11}\multirow{3}{*}{A2}      &CatRobot1&-2.54&-1.53&-2.14&-1.36&-1.95&-1.62&-1.53&-1.72&-1.32 \\
    \cline{2-11}                         &DaylightRoad2&-2.48&-2.06&-2.86&-1.84&-2.55&-2.29&-2.30&-1.69&-2.02 \\
    \cline{2-11}                         &ParkRunning3&-1.07&-0.78&-0.37&-0.41&-0.41&-0.43&-0.48&-0.36&-0.35 \\
    \cline{1-11}\multirow{5}{*}{B}       &MarketPlace&-1.88&-1.31&-1.99&-0.60&-0.55&0.41&-1.11&-1.61&0.88 \\
    \cline{2-11}                         &RitualDance&-3.48&-4.15&-3.44&-0.72&-0.69&-1.20&-1.04&-0.86&-1.53 \\
    \cline{2-11}                         &Cactus&-2.27&-1.61&-1.05&-1.08&-0.94&-0.37&-1.63&-1.35&-2.01\\
    \cline{2-11}                         &BasketballDrive&-2.35&-1.94&-2.38&-0.78&-1.39&-0.24&-1.10&-0.29&-1.06 \\
    \cline{2-11}                         &BQTerrace&-1.74&-1.63&-0.80&-0.78&-1.46&-1.23&-1.30&-1.12&-0.97 \\
    \cline{1-11}\multirow{4}{*}{C}       &BasketballDrill&-1.31&-2.29&-1.05&-0.41&-1.03&0.06&-1.22&-2.00&-1.41 \\
    \cline{2-11}                         &BQMall&-2.11&-0.34&-2.58&-1.14&-2.15&-1.16&-1.34&0.04&-0.12 \\
    \cline{2-11}                         &PartyScene&-1.50&-1.10&-2.58&-0.77&-0.57&-1.27&-0.90&-1.08&-0.52 \\
    \cline{2-11}                         &RaceHorsesC&-1.47&0.18&-0.42&-0.25&-1.02&-0.67&-0.65&-1.23&0.11 \\
    \cline{1-11}\multirow{4}{*}{D}       &BasketballPass&-0.47&-1.94&-3.35&-0.25&-1.25&0.70&0.18&-2.36&-0.67 \\
    \cline{2-11}                         &BQSquare&-1.26&-2.13&0.27&-0.55&-2.42&-3.05&-0.74&-0.07&-0.96 \\
    \cline{2-11}                         &BlowingBubbles&-1.11&-1.05&0.20&-0.32&-0.39&0.71&-1.03&-0.21&-0.78 \\
    \cline{2-11}                         &RaceHorses&-1.12&-1.04&0.81&-0.25&-1.01&0.40&-0.38&2.08&0.16 \\
    \cline{1-11}\multirow{3}{*}{E}       &FourPeople&-3.02&-1.73&-2.68&-2.04&-2.66&-2.93&-2.14&-1.28&-2.18 \\
    \cline{2-11}                         &Johnny&-1.88&-3.26&-2.15&-1.23&-0.35&-2.26&-1.06&-1.87&-0.71 \\
    \cline{2-11}                         &KristenAndSara&-2.22&-4.38&-2.66&-1.64&0.44&-2.44&-1.50&-1.16&-1.95 \\
    \cline{1-11}
    \multicolumn{2}{|c|}{{AVERAGE}}&-1.91&-1.73&-1.68&-0.87&-1.16&-0.91&-1.15&-0.95&-0.90\\
    \hline
   \end{tabular}
\end{center}
\end{table*}

\subsection{Coding Performance under the Latest VVC Test Model and Other Configurations}
In addition, the proposed method is evaluated on the platform of the latest VVC test model, i.e., VTM 16.0, in which DLIMD has been implemented. Besides AI configuration, the coding experiments are also conducted under \textbf{Low Delay P} (\textbf{LDP}) and \textbf{Random Access} (\textbf{RA}) configurations. It should be noted that the neural network is not changed after training with the instructions in Section 3.3.

The experimental results are shown in Table \ref{table10}, where the anchor is the original VTM 16.0 to calculate the value of BD-BR. It can be observed that the proposed method achieves 1.91\%, 0.87\% and 1.15\% bit rate reductions for Y component under AI, LDP and RA configurations, respectively. The coding gains are a little worse than those under VTM 5.0 shown in Table \ref{table5}. The reasons are that the neural network is not retrained, and the module of intra coding from VTM 5.0 to VTM 16.0 has been optimized.

\section{Conclusions}
In this paper, a deep learning based intra mode derivation method is presented to skip the module of intra mode signaling for saving coding bits. Instead of checking the candidate intra modes one by one to achieve the optimal, this process is casted into a multi-class classification task from signal processing to artificial intelligence. To adapt to variable coding blocks and different QP settings with one single model, the architecture is effectively developed. In particular, the hand-crafted and learned features are combined to compensate their individual limitations. The rate-distortion optimization is performed between the proposed method and the traditional method with a strategy flag signaled for performance competition. Compared with the state-of-the-art works, the proposed method achieves significant coding gains.


\bibliographystyle{ACM-Reference-Format}
\bibliography{sample-base}


\end{document}